\documentclass[a4paper,11pt]{article}
\pdfoutput=1
\usepackage{graphicx}
\usepackage{fullpage}
\usepackage{jheppub} % for details on the use of the package, please
\usepackage{bbm}
\usepackage{amsfonts}
\usepackage{slashed}
\usepackage{subfigure}
\usepackage{array}
\usepackage{verbatim}
\usepackage{booktabs}
\usepackage{color, soul}
\usepackage{mathrsfs}
\usepackage{epsfig}
\usepackage{graphicx}% Include figure files
\usepackage{dcolumn}% Align table columns on decimal point
\usepackage{bm}% bold math
\usepackage{amsmath}
\usepackage{amssymb}
\usepackage{multirow}

\setstcolor{red}
\let\jnfont=\rm
\def\NPB#1,{{\jnfont Nucl.\ Phys.\ B }{\bf #1},}
\def\PLB#1,{{\jnfont Phys.\ Lett.\ B }{\bf #1},}
\def\EPJC#1,{{\jnfont Eur.\ Phys.\ Jour.\ C }{\bf #1},}
\def\PRD#1,{{\jnfont Phys.\ Rev.\ D }{\bf #1},}
\def\PRL#1,{{\jnfont Phys.\ Rev.\ Lett.\ }{\bf #1},}
\def\MPLA#1,{{\jnfont Mod.\ Phys.\ Lett.\ A }{\bf #1},}
\def\JPG#1,{{\jnfont J.\ Phys.\ G}{\bf #1},}
\def\CTP#1,{{\jnfont Commun.\ Theor.\ Phys.\ }{\bf #1},}
\def\ZPC#1,{{\jnfont Z.\ Phys.\ C }{\bf #1},}
\def\JHEP#1,{{\jnfont JHEP \ }{\bf #1},}

\title{Strong constraints of LUX-2016 results on the natural NMSSM}

\author{Junjie Cao$^{1,2}$, Yangle He$^1$, Liangliang Shang$^1$, Wei Su$^3$, Peiwen Wu$^4$, Yang Zhang$^3$}

\affiliation{ $^1$  College of Physics and Materials Science,
        Henan Normal University, Xinxiang 453007, China \\
 $^2$ Department of Applied Physics, Xi'an Jiaotong University, Xi'an 710049, China \\
  $^3$ State Key Laboratory of Theoretical Physics,
      Institute of Theoretical Physics, Academia Sinica, Beijing 100190,
      China\\
  $^4$ School of Physics, KIAS, 85 Hoegiro, Seoul 02455, Republic of Korea }

\emailAdd{junjiec@itp.ac.cn}
\emailAdd{heyangle90@gmail.com}
\emailAdd{shlwell1988@gmail.com}
\emailAdd{weisv@itp.ac.cn}
\emailAdd{pwwu@kias.re.kr}
\emailAdd{zhangyang@itp.ac.cn}

\abstract{Given the fact that the relatively light Higgsino mass $\mu$ favored in natural supersymmetry usually results in a sizable scattering cross section between the neutralino
dark matter and the nucleon, we study the impact of the recently updated direct detection bounds from LUX experiment, including both Spin Independent (SI) and Spin Dependent (SD)
measurements, on the parameter space of natural Next-to-Minimal Supersymmetric Standard Model (nNMSSM). Different from the common impression that the SI bound is stronger than the SD
one, we find that the SD bound is complementary to the SI bound and in some cases much more powerful than the latter in limiting the nNMSSM scenarios.
After considering the LUX results, nNMSSM is severely limited, e.g. for the peculiar scenarios of the NMSSM where the next-to-lightest CP-even Higgs corresponds to the $125 {\rm GeV}$
Higgs boson discovered at the LHC,  the samples obtained in our random scan are excluded by more than $85\%$.  By contrast, the monojet search at the LHC Run-I can not exclude
any sample of nNMSSM. We also investigate the current status of nNMSSM and conclude that, although the parameter points with low fine tuning are still attainable,
they are distributed in some isolated parameter islands which are difficult to get.  Future dark matter direct search experiments such as XENON-1T
will provide a better test of nNMSSM.}

\begin{document}
\maketitle \indent
\newpage

\section{\label{introduction}Introduction}
From various cosmological and astrophysical observations, it has been a well established fact that over $20\%$ of the energy density of the Universe today is composed of Dark Matter (DM) \cite{PlanckCollaboration2015a}. Among the candidates predicted in new physics models beyond the Standard Model (SM), the Weakly Interacting Massive Particle (WIMP) is a very promising one which has a mass around the electroweak scale and couplings to the SM particles  close to the electroweak strength. As a typical example, the Lightest Supersymmetric Particle (LSP) in various Supersymmetric (SUSY) models falls into this category \cite{Jungman1995}. Motivated by the interactions predicted between DM and SM sector, many direct detection (DD) experiments are on going to search for possible scattering signals of DM particle off the nuclei.

Recently there are updated results from several groups including PICO-2L \cite{Amole2016}, PandaX-II \cite{Collaboration2016} and LUX \cite{LUXCollaboration2016-SI,LUXCollaboration2016}, and the results cover both Spin-Independent (SI) and Spin-Dependent (SD) scattering between DM and nuclei. In many cases the SI scattering is considered to be more promising in detecting DM signals due to its coherent property, of which the scattering cross section is proportional to $A^2$ of the nucleus and can benefit from the heavy nuclear elements \cite{Jungman1995}.
On the contrary, the SD scattering cross section suffers from the cancelation of the spins of nucleon pairs in the nucleus and thus does not have the $A^2$ enhancement  \cite{Belanger2008}. However, there are SUSY parameter space inducing cancelations in the SI amplitude and resulting in small and even vanishing SI cross section, the so-called Blind Spot (BS) \cite{Huang2014,Cheung2012,Badziak2015,Badziak:2016qwg}, in which case one has to consider the SD detection. Moreover, nuclei isotopes with un-paired nucleon and high abundance can be good targets to detect SD scattering, e.g. xenon used in XENON and LUX experiments and fluorine used in PICO experiments. Consequently, SI and SD detection methods are complementary to each other and should be considered together if one wants to constrain the parameter space of a certain model.

A particularly interesting SUSY scenario sensitive to DD experiment is natural SUSY (NS) \cite{Ellwanger2011,Baer2012,King:2012tr,Baer2013,Farina2013,Cao2014,Kim:2015dpa,Cao2016,Kim:2016rsd,Crivellin:2015bva}.
In this scenario, the Higgsino mass $\mu$ tends to be small, and consequently the lightest neutralino $\widetilde{\chi}^0_1$ as the DM candidate
contains sizable Higgsino components, which enables it to couple rather strongly with CP-even Higgs bosons $h_i$ and $Z$-boson. Given the fact that $t$-channel
exchange of $h_i$  ($Z$ boson) is the dominant contribution to SI (SD) cross section for DM-nucleon scattering
at tree-level, it is speculated that the continuously improved sensitivity in DD experiments can be promising
to test NS if $\widetilde{\chi}^0_1$ is fully responsible for the current
DM relic density. Since the NS scenario in the Minimal Supersymmetric Standard Model (MSSM) is theoretically unsatisfactory \cite{Cao2016}, we here investigate
this issue in the NS scenario of the Next-to-Minimal Supersymmetric Standard Model (NMSSM) \cite{Ellwanger2009}. To be more specific, we study the samples
obtained in \cite{Cao2016} which predict rather low fine tunings in getting electroweak observables $m_Z$ and $m_h$ and meanwhile satisfy various experimental
constraints, e.g. the DM relic density measured by WMAP and Planck \cite{WMAP,Planck}, the LUX-2015 limit on SI cross section
\cite{LUX-2015} as well as the direct searches for supersymmetric particles at LHC Run-I. Our analyses indicate that the constraints
from the LUX experiment in 2016 (LUX-2016), especially those from the upper bounds on SD cross section, are very strong in limiting
the NS scenario. Numerically speaking, we find that the LUX-2016 results can exclude more than $90\%$ Type III and Type IV samples
obtained in \cite{Cao2016}. Although the exact percentage may vary with different scan strategies, this number can exhibit the high sensitivity of the nNMSSM scenarios to the DD experiment. We note that the importance of the
SD cross section in limiting SUSY parameter space was not exhibited sufficiently before.

This paper is organized as follows. In Section \ref{Section-Basics}, we recapitulate the basics of the NMSSM including the calculation of the SI and SD scattering rates.
In Section \ref{Section-Strong} we illustrate
the capability of the LUX-2016 results of limiting the NS scenario of NMSSM (nNMSSM). We also study the status of nNMSSM
after the LUX-2016 experiment, which is presented in Section \ref{Section-Status}. Finally, we draw our conclusions in Section \ref{Section-Conclusion}.

\section{\label{Section-Basics}Basics about the NMSSM}

 In this section we briefly recapitulate the basics of the NMSSM including its Higgs and neutralino sectors, the naturalness argument and
 the calculation of SI and SD scattering rates. More detailed discussion and complete formulae about the basics can be found in \cite{Ellwanger2009,Cao2016,Badziak2015,Belanger2008} and references therein.

\subsection{Natural NMSSM}
The superpotential of the $Z_3$-invariant NMSSM takes the following form \cite{Ellwanger2009}
\begin{eqnarray}
  W_{\rm NMSSM} &=& W_F + \lambda\hat{H_u} \cdot \hat{H_d} \hat{S}
  +\frac{1}{3}\kappa \hat{S^3},
 \end{eqnarray}
where $W_F$ is the superpotential of the MSSM without the $\mu$-term, and $\lambda$, $\kappa$ are dimensionless parameters
describing the interactions among the Higgs superfields.
The Higgs potential of the NMSSM consists of the F-term and D-term of the superfields, as well as the soft breaking terms
\begin{eqnarray}
V_{\rm NMSSM}^{\rm soft} &=& m_{H_u}^2 |H_u|^2 + m_{H_d}^2|H_d|^2
  + m_S^2|S|^2 \nonumber \\
  &+& ( \lambda A_{\lambda} SH_u\cdot H_d
  +\frac{1}{3}\kappa A_{\kappa} S^3 + h.c.), \label{input-parameter1}
\end{eqnarray}
where $H_u$, $H_d$ and $S$ denote the scalar component of the superfields $\hat{H}_u$,  $\hat{H}_d$ and $\hat{S}$,
respectively. In practice, it is convenient to rotate the fields $H_u$ and $H_d$ by
\begin{eqnarray}
H_1=\cos\beta H_u + \varepsilon \sin\beta H_d^*, ~~
H_2=\sin\beta H_u - \varepsilon \cos\beta H_d^*, ~~H_3 = S,
\end{eqnarray}
where $\varepsilon$ is an antisymmetric tensor with $\varepsilon_{12}=-\varepsilon_{21}=1$ and $\varepsilon_{11}=\varepsilon_{22}=0$,
 and $\tan \beta \equiv v_u/v_d$ with $v_u$ and $v_d$ representing
the vacuum expectation value of $H_u$ and $H_d$ fields, respectively. After this rotation, the redefined fields $H_i$ ($i=1,2,3$) have the following form
\begin{eqnarray}
H_1 = \left ( \begin{array}{c} H^+ \\
       \frac{S_1 + i P_1}{\sqrt{2}}
        \end{array} \right),~~
H_2 & =& \left ( \begin{array}{c} G^+
            \\ v + \frac{ S_2 + i G^0}{\sqrt{2}}
            \end{array} \right),~~
H_3  = v_s +\frac{1}{\sqrt{2}} \left(  S_3 + i P_2 \right).
\label{fields}
\end{eqnarray}
where $H_2$ corresponds to the SM Higgs doublet with $G^+, G^0$ being the Goldstone bosons eaten by $W$ and $Z$ bosons
respectively, while $H_1$ represents a new $SU(2)_L$ doublet scalar field
and it has no coupling to $W$ and $Z$ bosons at tree-level. From Eq.(\ref{fields}), it is obvious that the Higgs sector
of the NMSSM consists of three CP-even mass eigenstates $h_1$, $h_2$ and $h_3$, which are the mixtures of $S_1$, $S_2$ and $S_3$,
two CP-odd mass eigenstates $A_1$ and $A_2$ composed by $P_1$ and $P_2$, as well as two charged Higgs $H^\pm$.
In the following, we assume $m_{h_1} < m_{h_2} < m_{h_3}$ and $m_{A_1} < m_{A_2}$, and call $h_i$ the SM-like Higgs boson
if its dominant component is composed of the field $S_2$. The eigenstates $h_i$ are related to the fields $S_j$ by
\begin{eqnarray}
h_i = \sum_{j=1}^3 U_{ij} S_j,
\end{eqnarray}
with $U$ being the rotation matrix to diagonalize the mass matrix for the $S_i$ fields.

An interesting feature of NMSSM is that the squared mass term of the filed $S_2$ in the SM-like Higgs double $H_2$ is given by
\begin{eqnarray}
m_{S_2 S_2}^2 = m_Z^2 \cos^2 2 \beta + \lambda^2 v^2  \sin^2 2 \beta, \nonumber
\end{eqnarray}
where the first term on the right side is the MSSM contribution, and the second term is peculiar to any gauge singlet extension of the MSSM.
Moreover, if the relation $m_{S_3 S_3}^2 < m_{S_2 S_2}^2$ holds, the mixing between the fields $S_2$ and $S_3$ can further enhance the SM-like
Higgs mass. In this case, $h_1$ is a singlet-dominate scalar and $h_2$ plays the role of the SM Higgs boson.
Benefiting from these features, $m_{h_2} \simeq 125 {\rm GeV}$ does not necessarily require a large radiative contribution from stop loops \cite{Cao2012,Ellwanger2011-diphoton}.

Instead of using the soft parameters $m_{H_u}^2$, $m_{H_d}^2$ and $m_S^2$, one usually trades them for $m_Z$, $\tan \beta$ and $\mu \equiv \lambda v_s $ by implementing the scalar potential minimization conditions, and uses the following input parameters
\begin{eqnarray}
\lambda, \quad \kappa, \quad \tan \beta, \quad \mu, \quad M_A, \quad A_\kappa,  \label{input-parameter}
\end{eqnarray}
where the parameter $A_\lambda$ in Eq.(\ref{input-parameter1}) is replaced by the squared mass of the CP-odd field $P_1$ given by
\begin{eqnarray}
M^2_A \equiv m_{P_1 P_1}^2 = \frac{2\mu}{\sin2\beta}(A_{\lambda}+\kappa v_s).
\end{eqnarray}
Note that $M_A$ represents the mass scale of the doublet $H_1$ and is usually larger than about $300 \,{\rm GeV}$
from the LHC searches for non-standard doublet Higgs bosons.

The neutralino sector of the  NMSSM consists of the fields Bino $\tilde{B}^0$, Wino $\tilde{W}^0$, Higgsinos $\tilde{H}_{d,u}^0$ and Singlino $\tilde{S}^0$,
which is the fermion component of the superfield $\hat{S}$. Taking the basis $\psi^0 = (-i \tilde{B}^0, - i \tilde{W}^0, \tilde{H}_{d}^0, \tilde{H}_{u}^0,
\tilde{S}^0)$, one has the following symmetric neutralino mass matrix
\begin{equation}
{\cal M} = \left(
\begin{array}{ccccc}
M_1 & 0 & -\frac{g_1 v_d}{\sqrt{2}} & \frac{g_1 v_u}{\sqrt{2}} & 0 \\
  & M_2 & \frac{g_2 v_d}{\sqrt{2}} & - \frac{g_2 v_u}{\sqrt{2}} &0 \\
& & 0 & -\mu & -\lambda v_u \\
& & & 0 & -\lambda v_d\\
& & & & \frac{2 \kappa}{\lambda} \mu
\end{array}
\right). \label{eq:MN}
\end{equation}
where $M_1$ and $M_2$ are Bino and Wino soft breaking mass respectively. In the limit of $|M_1|, |M_2| \gg |\mu|$,
the Bino and Wino components decouple from the mixing, and the remaining three light
neutralinos $\tilde{\chi}_i^0$ ($i=1,2,3$) can be decomposed into
\begin{eqnarray}
\tilde{\chi}_i^0 & \approx & N_{i3} \tilde{H}_d^0 + N_{i4}\tilde{H}_u^0 + N_{i5} \tilde{S}^0, \label{neutralino-mass}
\end{eqnarray}
where the elements of the rotation matrix $N$ can be approximated by
\begin{eqnarray}
N_{i3}:N_{i4}:N_{i5} \simeq \lambda (v_d \mu -  v_u m_{\tilde{\chi}_i^0}):
\lambda (v_u \mu -  v_d m_{\tilde{\chi}_i^0}): (m_{\tilde{\chi}_i^0}^2 - \mu^2)  \label{neutralino-mixing}
\end{eqnarray}
with $m_{\tilde{\chi}_i^0}$ denoting the mass of $\tilde{\chi}_i^0$. In this case, the DM candidate may be
either Singlino-dominated or Higgsino-dominated \cite{Cao2016}. As has been pointed out by numerous studies, DM may achieve its measured relic
density in following regions
\begin{itemize}
\item Higgs boson and $Z$ boson resonance regions, where the Higgs boson may be any of the three CP-even and two
CP-odd Higgs bosons.
\item Coannihilation region where $\tilde{\chi}_1^0$ is nearly degenerated with any of $\tilde{\chi}_1^\pm$, $\tilde{\chi}_2^0$ and $\tilde{l}$ 
($\tilde{l}$ represents the lightest slepton).
\item Region in which $\tilde{\chi}_1^0$ has moderate Higgsino and Singlino fractions.
\end{itemize}

Naturalness in the NMSSM can be measured by the following two quantities \cite{Baer2013}
\begin{eqnarray}
\Delta_Z = \max_i |\frac{\partial \log m_Z^2}{\partial \log p_i}|, \quad \Delta_h = \max_i |\frac{\partial \log m_h^2}{\partial \log p_i}|.
\end{eqnarray}
Here $h$ denotes the SM-like Higgs boson, and $p_i$ are parameters defined at the weak scale including those in Eq.(\ref{input-parameter}) and the
top quark Yukawa coupling $Y_t$ which is used to estimate the sensitivities of $m_Z$ and $m_h$ to stop masses \cite{Ellwanger2011}. Apparently,
$\Delta_Z$ ($\Delta_h$) reflects the sensitivity of $m_Z$ ($m_h$) to SUSY parameters at weak scale
and a larger value for any of  $\Delta_Z$ and $\Delta_h$ corresponds to more tuning. Formulae of calculating $\Delta_Z$ and  $\Delta_h$ can be found in \cite{Ellwanger2011} and \cite{Farina2013} respectively.

 \subsection{Blind spot in spin independent cross section}

In the NMSSM with heavy squarks, the dominant contribution to SI DM-nucleon scattering comes from $t$-channel exchange of the CP-even Higgs bosons
\cite{Drees1993,Drees1992,Jungman1995,Belanger2008}. The SI cross section is then expressed as
\begin{eqnarray}
\sigma^{SI}_{\tilde{\chi}-(n)} = \frac{4 \mu_r^2}{\pi} |f^{(n)}|^2 , \label{SI-cross}
\end{eqnarray}
where $n=\{p,n\}$ denotes nucleon, $\mu_r$ is the reduced mass of DM and the nucleon, and\footnote{We remind that in SUSY, the SI cross sections for DM-proton scattering and DM-neutron scattering are roughly equal, i.e.
$\sigma^{SI}_{\tilde{\chi}-p} \simeq \sigma^{SI}_{\tilde{\chi}-n}$ (see for example \cite{DM-detecion-SI-SD}).}
\begin{eqnarray}
f^{(n)} \approx \sum^3_{i=1} f^{(n)}_{h_i} = \sum^3_{i=1} \frac{C_{h_i \chi \chi} C_{h_i n n}}{2m^2_{h_i}},
\end{eqnarray}
with $C_{h_i \chi \chi}$ ($C_{h_i n n}$) representing the coupling of $h_i$ with DM (nucleon) \cite{Badziak2015,Badziak:2016qwg}. The explicit expressions of
$C_{h_i \chi \chi}$ and $C_{h_i n n}$ are given by \cite{Ellwanger2009}
\begin{eqnarray}
\alpha_{h_i\chi\chi}
\!\!&=&\!\!
\sqrt{2}\lambda
\left(U_{i2}N_{14}N_{15}+U_{i1}N_{13}N_{15}+S_{i3}N_{13}N_{14}\right)
-\sqrt{2}\kappa U_{i3}N_{15}^2
\nonumber  \\
\!\!&+&\!\!
g_1\left(U_{i2}N_{11}N_{13}-U_{i1}N_{11}N_{14}\right)
-g_2\left(U_{i2}N_{12}N_{13}-U_{i1}N_{12}N_{14}\right), \\
\alpha_{h_inn}
&=& \frac{m_n}{\sqrt{2}v}
\left( \frac{U_{i2}}{\cos\beta}F^{(n)}_d +\frac{U_{i1}}{\sin\beta}F^{(n)}_u
\right)\,,
\end{eqnarray}
where $v^2=v_u^2+v_d^2\approx(174 \, \rm{GeV})^2$, $F^{(n)}_d=f^{(n)}_d+f^{(n)}_s+\frac{2}{27}f^{(n)}_G$
and $F^{(n)}_u=f^{(n)}_u+\frac{4}{27}f^{(n)}_G$ with $f^{(n)}_q =m_N^{-1}\left<n|m_qq\bar{q}|n\right> $
for $q=u,d,s$ denoting the normalized light quark contribution to nucleon mass, and $f^{(n)}_G=1-\sum_{q=u,d,s}f^{(n)}_q$
related with other heavy quark mass fraction in nucleon \cite{Drees1993,Drees1992}.

The blind spot is defined as the SUSY parameter point for which the SI cross-section vanishes.
From the formula in Eq.(\ref{SI-cross}), one can get the analytic expression of the BS condition which was studied in detail in
\cite{Badziak2015,Badziak:2016qwg}. This condition correlates the parameters $\lambda$, $\kappa$, $\tan \beta$, $\mu$, $m_A$ and
$m_{\tilde{\chi}_1^0}$ in a nontrivial way and in some special cases its expression is quite simple.
In the following, we consider two specific situations to illustrate this point.
\begin{itemize}
\item Case I: $\tan \beta \gg 1$,  and both the gaugino fields and the singlet field $S_3$ decouple from the DM-nucleon scattering. In this case,
the BS condition takes the simple form  (see Eq.(50) in \cite{Badziak2015})
\begin{eqnarray}
\frac{m_{\chi}}{\mu}-\sin2\beta
\approx \left(\frac{m_h}{m_H}\right)^2\frac{\tan\beta}{2}.
\end{eqnarray}
\item Case II: $h_1$ and $h_2$ correspond to the singlet-dominated and SM-like Higgs boson respectively, and both the gaugino fields and
the heavy doublet field $S_1$ decouple from the DM-nucleon scattering. For this case, the BS condition reads (see Eq.(64) in \cite{Badziak2015})
\begin{eqnarray}
\frac{m_{\chi}}{\mu}-\sin2\beta
\approx
-\frac{\gamma+\mathcal{A}_s}{1-\gamma\mathcal{A}_s}
\eta^{-1}\left(\frac{m_{\chi}}{\mu}-\sin2\beta\right)\,,
\end{eqnarray}
with $\gamma$, $\eta$ and $\mathcal{A}_s$ defined by
\begin{eqnarray}
\gamma &=& \frac{U_{23}}{U_{22}}, \quad \eta \equiv
\frac{N_{15}(N_{13}\sin\beta+N_{14}\cos\beta)}
{N_{13}N_{14}-\frac{\kappa}{\lambda}N_{15}^2}, \nonumber \\
\mathcal{A}_s &\approx &
-\gamma\frac{1+c_1}{1+c_2}\left(\frac{m_{h_2}}{m_{h_1}}\right)^2
\,, \quad  c_i = 1+\frac{U_{i1}}{U_{i2}}\left(\tan\beta- \cot \beta \right). \nonumber
\end{eqnarray}
\end{itemize}

We remind that the LUX-2016 experiment has imposed an upper bound of the SI cross section at the order
of $10^{-45} {\rm cm}^2$. Confronted with such a situation, one can infer that
the cancelation among the different $h_i$ contributions usually exists, and if no signal is detected
in future DD experiments BS will become important for SUSY
to coincide with experimental results.

%%%%%%%%Fig.1%%%%%%%%%%%%%%%%%%%%%%%%%%%%%%%%%%%%%%%%%%%%%%%%%%%%%%%%%%%%
\begin{figure}[t]
\centering
\includegraphics[height=8cm,width=16cm]{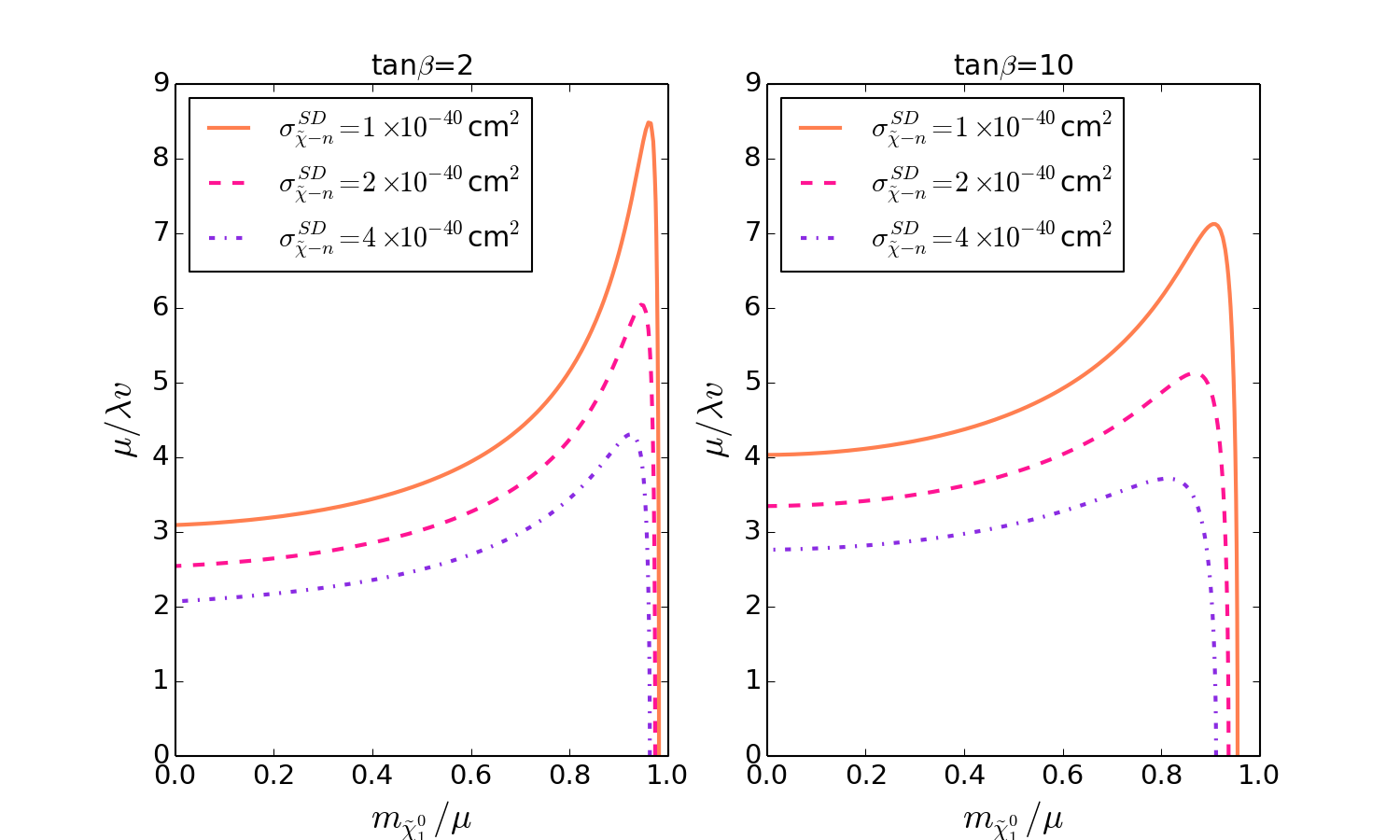} \\
\caption{Constant contours of the SD cross section for DM-neutron scattering projected on
$\mu/(\lambda v)-m_{\tilde{\chi}_1^0}/\mu$ plane for $\tan \beta =2$ (left panel) and  $\tan \beta =10$ (right panel).
This figure indicates that the upper bound of the LUX experiment on  $\sigma^{SD}_{\tilde{\chi}-n}$
is able to exclude the parameter region with $\mu/(\lambda v) \lesssim 3$, and the tightest limit comes from the region of
$m_{\tilde{\chi}_1^0}/\mu \sim (0.8-0.9)$.   \label{fig1}}
\end{figure}

\subsection{Spin dependent cross section}

In the heavy squark limit, only $t$-channel $Z$ exchange diagram contributes to the SD cross section at tree level in the NMSSM. The cross section
is then given by \cite{Badziak2015,Badziak:2016qwg}
\begin{eqnarray}
\sigma_{\tilde{\chi}-(n)}^{SD} \simeq C^{(n)} \times 10^{-4} \, { \rm{pb} } \,\bigg( \frac{|N_{13}|^2 - |N_{14}|^2}{0.1} \bigg)^2 \label{SD-Expression-1}
\end{eqnarray}
with $C^{p} \approx 4.0$ and $C^{n} \approx 3.1$ for the typical values of $f^{(n)}_q$. With the decoupling of gauginos, we have the following simple expression
\begin{eqnarray}
N_{13}^2-N_{14}^2 & \simeq & \frac{[1-(m_{\tilde{\chi}_1^0}/\mu)^2] (1-N^2_{15}) \cos 2 \beta}{1+(m_{\tilde{\chi}_1^0}/\mu)^2-2(m_{\tilde{\chi}_1^0}/\mu)\sin 2\beta}
\nonumber \\
&\simeq  &\frac{\left[1-\left(m_{\tilde{\chi}_1^0}/\mu\right)^2\right]
\cos2\beta}{1+\left(m_{\tilde{\chi}_1^0}/\mu\right)^2-2\left(m_{\tilde{\chi}_1^0}/\mu\right)\sin2\beta
+\left[1-\left(m_{\tilde{\chi}_1^0}/\mu\right)^2\right]^2
\left({\mu}/{\lambda v}\right)^2}\, \label{SD-Expression-2}
\end{eqnarray}
by using the approximation in Eq.(\ref{neutralino-mixing}). From Eqs.(\ref{SD-Expression-1}) and (\ref{SD-Expression-2}), one can
immediately see that $\sigma^{SD}_{\tilde{\chi}-n} \simeq 0.76 \times \sigma^{SD}_{\tilde{\chi}-p}$, and  $\sigma^{SD}_{\tilde{\chi}-n}$
vanishes with $\tan\beta=1$ or pure Higgsino/Singlino DM. Given that the LUX-2016 limit on $\sigma^{SD}_{\tilde{\chi}-n}$ is much stronger
than that on $\sigma^{SD}_{\tilde{\chi}-p}$, we hereafter only consider $\sigma^{SD}_{\tilde{\chi}-n}$
in the following discussion \footnote{Note that although the PICO limit on $\sigma^{SD}_{\tilde{\chi}-p}$ is much stronger than
that of the LUX-2016 experiment \cite{Amole2016,LUXCollaboration2016}, it is still weaker than the LUX-2016 limit on  $\sigma^{SD}_{\tilde{\chi}-n}$ in constraining SUSY parameter space
after considering the correlation $\sigma^{SD}_{\tilde{\chi}-n} \simeq 0.76 \times \sigma^{SD}_{\tilde{\chi}-p}$.}. From the expressions one can also get constant contours of
$\sigma^{SD}_{\tilde{\chi}-n}$ on $\mu/(\lambda v)-m_{\tilde{\chi}_1^0}/\mu$ plane, which are shown in Fig.\ref{fig1} for $\tan \beta=2, 10$ respectively.
This figure indicates that for fixed $m_{\tilde{\chi}_1^0}/\mu$, the SD cross section decreases monotonously with the increase
of $\mu/(\lambda v)$, while for fixed $\mu/(\lambda v)$  the cross section increases
with the increase of $m_{\tilde{\chi}_1^0}/\mu$ to reach its maximum at $m_{\tilde{\chi}_1^0}/\mu \sim (0.8-0.9)$. Given that
the LUX-2016 limit on $\sigma^{SD}_{\tilde{\chi}-n}$ is at the order of  $10^{-40} {\rm cm}^2$ \cite{LUXCollaboration2016},
one can infer that the experiment can exclude the parameter region with $\mu/(\lambda v) \lesssim 3$ and
the tightest limit comes from the region $m_{\tilde{\chi}_1^0}/\mu \sim (0.8-0.9)$.

Throughout this work, we use the package NMSSMTools \cite{NMSSMTools} to get the particle spectrum of the NMSSM, the package micrOMEGAs \cite{micrOMEGA}
to calculate DM relic density and the SI and SD cross sections \footnote{We emphasize that the formulae for the BS conditions and the SD cross section
rely heavily on certain assumptions, and can not be applied to all the samples encountered in our study. So in calculation we use the exact expressions of
the cross sections implemented in micrOMEGAS \cite{micrOMEGA} to get their values. }. We use the default setting of micrOMEGAs, i.e.
$\sigma_{\pi N} = 34\,  {\rm MeV}$ and $\sigma_0 = 42\,  {\rm MeV}$, to get the values of $f^{(n)}_q$ \cite{Drees1993,Drees1992}.
 We checked that if we take $\sigma_{\pi N} = 59\,  {\rm MeV}$ from \cite{SI-piN} and $\sigma_0 = 58\,  {\rm MeV}$ from \cite{SI-pi0,SI-pi0-2,SI-pi0-3},
the SI cross section will be enhanced by a factor from $20\%$ to $40\%$. Other related discussions can be found in \cite{Crivellin:2013ipa,Hoferichter:2015dsa,Hoferichter:2016ocj}.

\section{\label{Section-Strong}Strong constraints of the LUX-2016 results on nNMSSM}

In order to study the constraints of the LUX-2016 results on nNMSSM, we consider the samples discussed in our previous work \cite{Cao2016}. These samples
were obtained in the following way:
\begin{itemize}
\item First, we fixed all soft breaking parameters for first two generation squarks and gluino mass at $2\, {\rm TeV}$. We also
assumed a common value for all soft breaking parameters in slepton sector (denoted by $m_{\tilde{l}}$ hereafter) and
$m_{U_3} = m_{D_3}$, $A_t = A_b$ for soft breaking parameters in third generation squark section.
\item Second, we scanned  by Markov Chain method following parameter
space of the NMSSM
\begin{eqnarray}\label{NMSSM-scan}
&& 0 <\lambda\leq 0.75,\quad  0 <\kappa \leq 0.75, \quad  2 \leq \tan{\beta} \leq 60,\quad  100{\rm ~GeV}\leq m_{\tilde{l}} \leq 1 {\rm ~TeV},  \nonumber \\
&& 100 {\rm GeV} \leq \mu \leq 1 {\rm TeV}, \quad 50 {\rm ~GeV}\leq M_A \leq 2 {\rm ~TeV}, \quad |A_{\kappa}| \leq 2 {\rm TeV}, \nonumber\\
&& 100{\rm ~GeV}\leq M_{Q_3},M_{U_3} \leq 2 {\rm ~TeV}, \quad  |A_{t}|\leq {\rm min}(3 \sqrt{M_{Q_3}^2 + M_{U_3}^2}, 5 {\rm TeV}), \nonumber\\
&& 20 {\rm GeV} \leq M_1 \leq 500 {\rm GeV}, \quad 100 {\rm GeV} \leq M_2 \leq 1 {\rm TeV}.
\end{eqnarray}
The likelihood function we adopted is $L= L_{\Omega} \times L_{m_h}$, where
\begin{eqnarray}
L_{\Omega} = \exp\left[-\frac{(\Omega_{\rm th} - \Omega_{\rm obs})^2}{2\,(\delta\Omega)^2}\right], \quad L_{m_h}  =  \exp\left[-\frac{(m_{th} - m_{obs})^2 }
	{2(\delta m_h)^2} \right ].
\end{eqnarray}
In above expressions, $\Omega_{\rm obs} = 0.1198/h^2$ with $h$ being the normalized Hubble constant is the cosmological DM parameter obtained in the latest PLANCK results\,\cite{Planck},
$\delta \Omega$ is the error including both the observational and theoretical uncertainties as $(\delta\Omega)^2 = (0.0015/h^2)^2 + (0.025\,\Omega_{\rm th})^2$ \cite{Likehood},
$m_{th}$ ($m_{obs} = 125.09\, {\rm GeV}$ \cite{Aad:2015zhl}) is the theoretical prediction (measured value) of the SM-like Higgs boson mass, and $\delta m_h = 3\, {\rm GeV}$ is its total (theoretical and experimental) uncertainty.

\item Third, we picked up physical samples from the scan by requiring them to satisfy $\Delta_Z \leq 50$, $\Delta_h \leq 50$ and
all the constraints contained in the package NMSSMTools-4.9.0 \cite{NMSSMTools}, such as various B-physics observables at $2\sigma$ level, the DM
relic density at $2\sigma$ level, and the LUX-2015 limit on SI scattering rate \cite{LUX-2015}. We also considered the limitations from the
direct searches for Higgs bosons at LEP, Tevatron and LHC by using the package HiggsBounds \cite{HiggsBounds} and
performed the $125\, {\rm GeV}$ Higgs data fit with the package  HiggsSignal \cite{HiggsSignals}. Moreover, we emphasize here that
the constraints from various searches for SUSY at LHC Run-I on the samples were also implemented by detailed  Monte Carlo simulations,
which is the core of our previous work \cite{Cao2016}.
\end{itemize}
According to the analysis in \cite{Cao2016}, the physical samples  can be classified into four types:
for Type I samples, $h_1$ corresponds to the SM-like Higgs boson and $\tilde{\chi}_1^0$ is Bino-dominated,
while for Type II, III and IV samples, $h_2$ acts as the SM-like Higgs boson with $\tilde{\chi}_1^0$
being Bino-, Singlino- and Higgsino-dominated respectively. Among the four types of the samples, the lowest
fine-tuning comes from Type III and Type IV samples, for which $\Delta_Z$ and $\Delta_h$ may be as low as
about 2 and therefore they are of particular interest to us.

%%%%%%%%Fig.2%%%%%%%%%%%%%%%%%%%%%%%%%%%%%%%%%%%%%%%%%%%%%%%%%%%%%%%%%%%%
\begin{figure}[t]
\centering
\includegraphics[height=7.2cm,width=7.2cm]{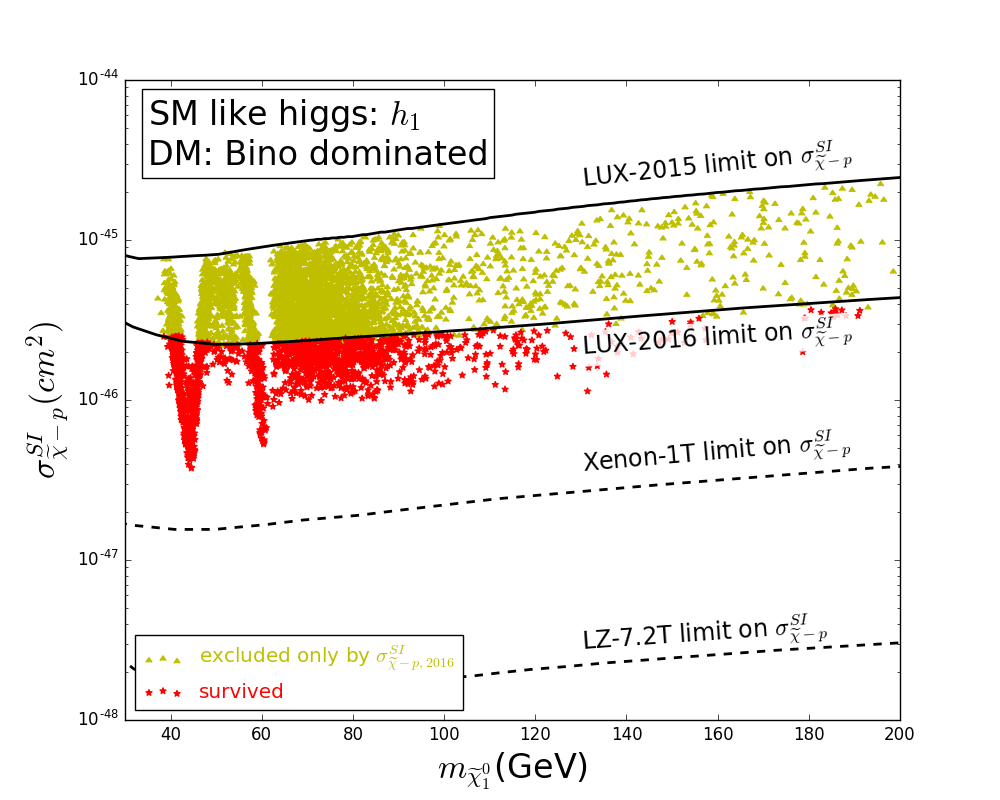} \hspace{-0.5cm}
\includegraphics[height=7.2cm,width=7.2cm]{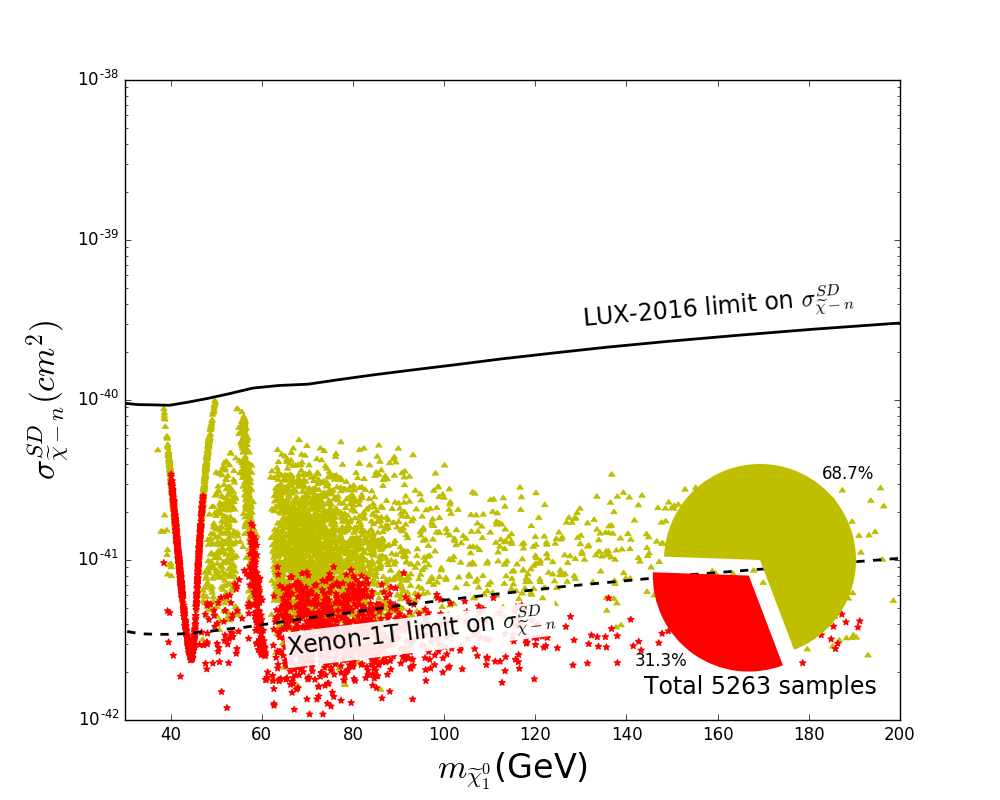}
\vspace{-0.4cm}
\caption{Type I samples projected on $\sigma_{\tilde{\chi}-p}^{SI}-m_{\tilde{\chi}_1^0}$ plane (left panel) and  $\sigma_{\tilde{\chi}-n}^{SD}-m_{\tilde{\chi}_1^0}$
plane (right panel) respectively. The dark green samples are excluded by the LUX-2016 limit on SI cross section,
while the red ones survive the limit. Note that the LUX-2016 limit on SD cross section for DM-neutron scattering
can not exclude any Type I samples. \label{fig2}}
\end{figure}
%%%%%%%%%%%%%%%%%%%%%%%%%%%%%%%%%%%%%%%%%%%%%%%%%%%%%

Now we show the impact of the LUX-2016 results on the samples. In Fig.\ref{fig2}, we project Type I samples on
$\sigma_{\tilde{\chi}-p}^{SI}-m_{\tilde{\chi}_1^0}$ plane (left panel) and  $\sigma_{\tilde{\chi}-n}^{SD}-m_{\tilde{\chi}_1^0}$
plane (right panel) respectively. The samples marked by dark green color are excluded by the LUX-2016 limit on SI cross
section, while those marked by red color survive the limit. From the figure one can learn that
the SI cross sections of the samples are usually larger than $3 \times 10^{-47} {\rm cm^2}$, which is
within the detection sensitivity of future XENON-1T experiment \cite{Cushman:2013zza}.
One can also learn that among the total 5263 Type I samples, $68.7 \%$ of them have been excluded by the
LUX-2016 limit on SI cross section, and by contrast the LUX-2016 limit on SD cross section for DM-neutron scattering can not exclude any
Type I samples. The underlying reason for these features
can be inferred from the formulae that \footnote{These formulae are valid in the case that $|M_2| \gg \mu, M_1$ and
$\lambda \rightarrow 0$ \cite{Cao:2015loa,Cao:2015efs}, i.e. the Wino and Singlino fields decouple from the neutralino mixing.}
\begin{eqnarray}
C_{h_1 \bar{\tilde{\chi}}_1^0 \tilde{\chi}_1^0} \propto e \frac{m_Z}{\mu} \Big[ \cos (\beta + \alpha) + \sin (\beta -\alpha) \frac{M_1}{\mu} \Big], \quad C_{Z \bar{\tilde{\chi}}_1^0 \tilde{\chi}_1^0} \propto e \frac{m_Z^2}{\mu^2} \cos 2 \beta ( 1 - \frac{M_1^2}{\mu^2} ).  \nonumber
\end{eqnarray}
So for moderate light $\mu$ which is required to predict a small $\Delta_Z$, the SI cross section is rather
large (in comparison with its LUX-2016 limit) given no strong cancelation between $h_1$ (the SM-like Higgs boson) contribution
and the other contributions. On the other hand, because the coupling between DM and $Z$ boson is proportional to
$m_Z^2/\mu^2 \cos 2 \beta$ and thus suppressed in comparison
with the $h_1 \bar{\tilde{\chi}}_1^0 \tilde{\chi}_1^0$ coupling, the SD cross section is not significant with respect to its
experimental limit. We remind that in the scan, we do not include any information about the
DD experiment in the likelihood function.
Although the exact percentage of excluded samples may vary with different scan strategies,
the number shown above can still reflect the powerfulness of the LUX-2016 results on Type I samples.

%%%%%%%%Fig.3%%%%%%%%%%%%%%%%%%%%%%%%%%%%%%%%%%%%%%%%%%%%%%%%%%%%%%%%%%%%
\begin{figure}[t]
\centering
\includegraphics[height=7.2cm,width=7.2cm]{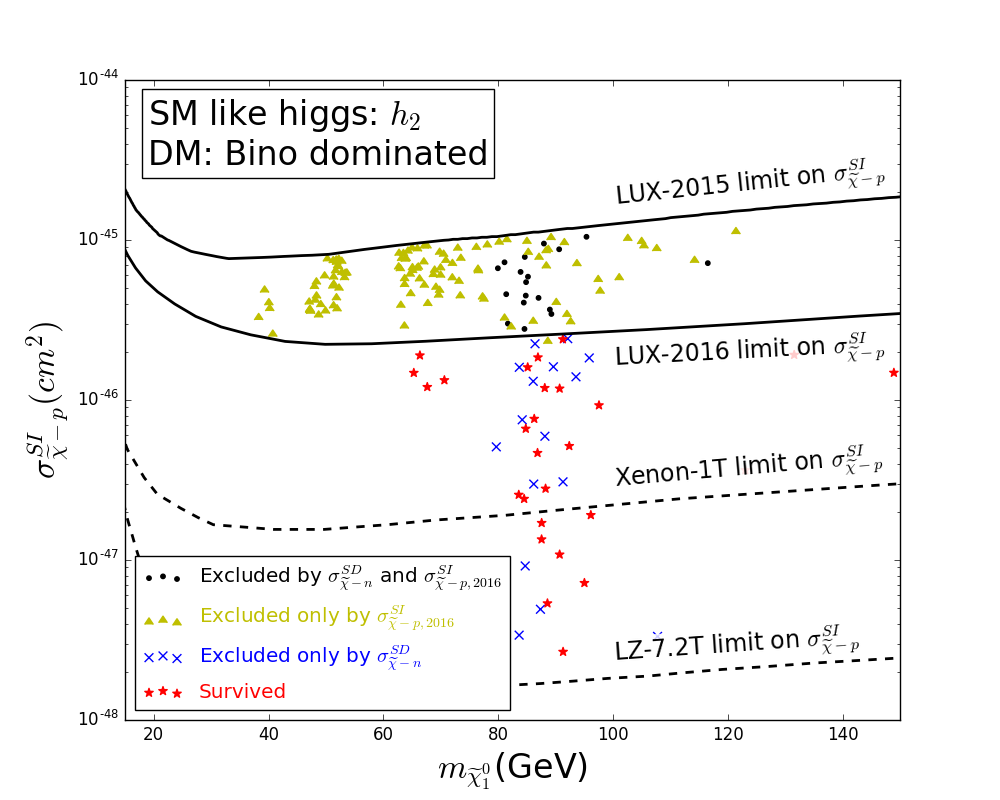} \hspace{-0.5cm}
\includegraphics[height=7.2cm,width=7.2cm]{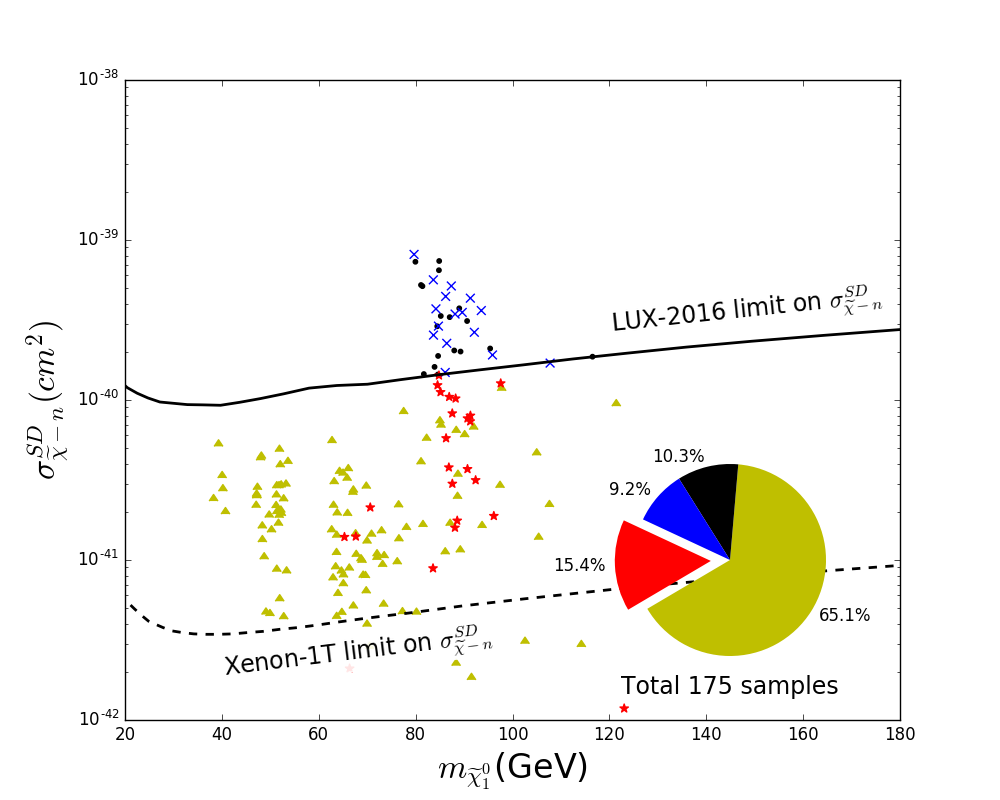}
\vspace{-0.4cm}
\caption{Type II samples projected on $\sigma_{\tilde{\chi}-p}^{SI}-m_{\tilde{\chi}_1^0}$ plane (left panel) and  $\sigma_{\tilde{\chi}-n}^{SD}-m_{\tilde{\chi}_1^0}$
plane (right panel) respectively. The black samples are excluded by both the LUX-2016 limit
on SI cross section for DM-proton scattering and that on SD cross section for DM-neutron scattering.
By contrast the dark green ones and the blue ones are excluded only by either of the limits,
and the red ones survive all the limits. The percentages of each kind of colored samples in the total Type II samples
are shown by the disc on the right panel.
\label{fig3}}
\end{figure}
%%%%%%%%%%%%%%%%%%%%%%%%%%%%%%%%%%%%%%%%%%%%%%%%%%%%%

%%%%%%%%Fig.4%%%%%%%%%%%%%%%%%%%%%%%%%%%%%%%%%%%%%%%%%%%%%%%%%%%%%%%%%%%%
\begin{figure}[t]
\centering
\includegraphics[height=7.2cm,width=7.2cm]{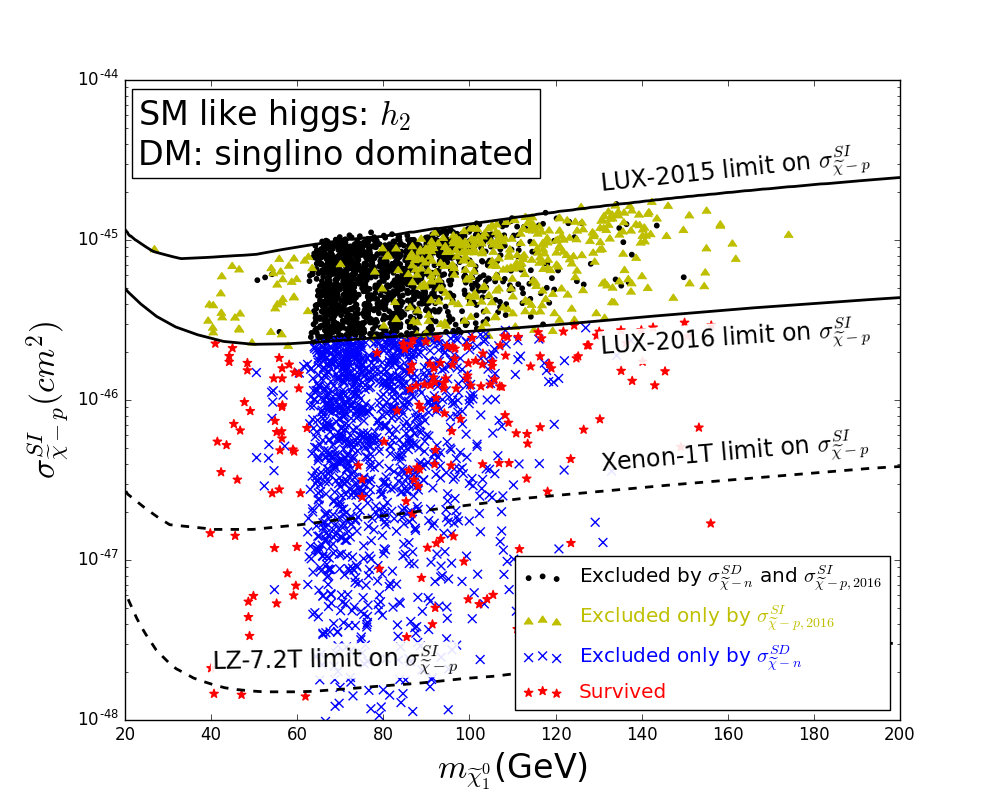} \hspace{-0.5cm}
\includegraphics[height=7.2cm,width=7.2cm]{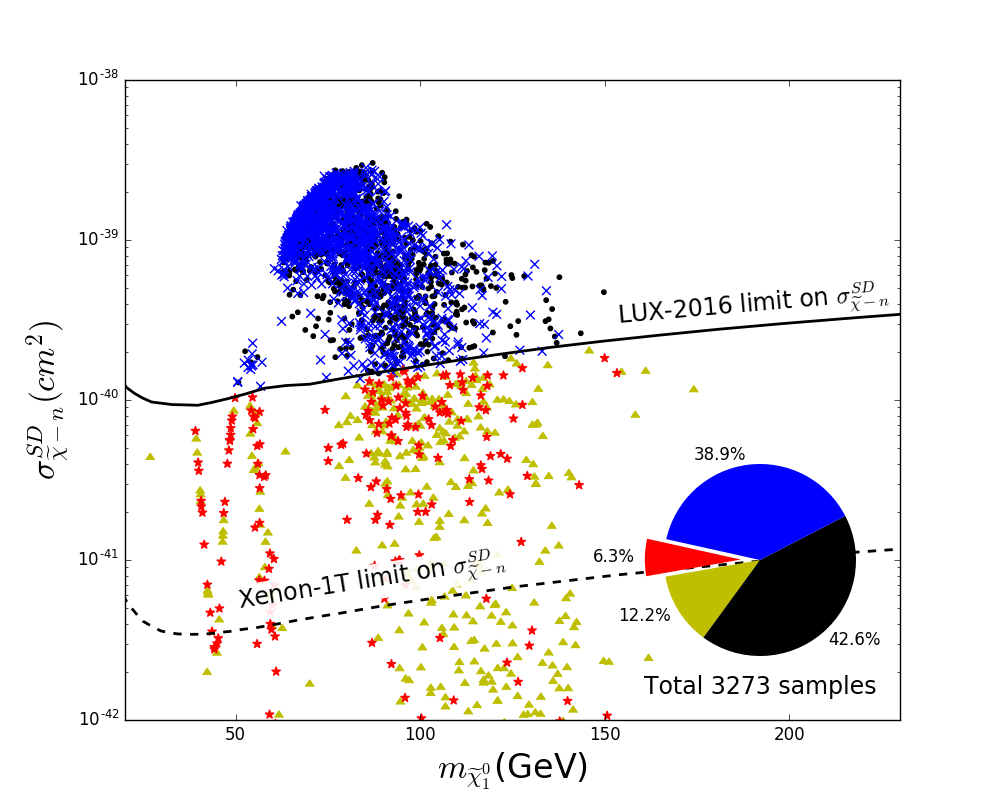}
\vspace{-0.4cm}
\caption{Same as Fig.\ref{fig3}, but for Type III samples.  \label{fig4}}
\end{figure}
%%%%%%%%%%%%%%%%%%%%%%%%%%%%%%%%%%%%%%%%%%%%%%%%%%%%%

%%%%%%%%Fig.4%%%%%%%%%%%%%%%%%%%%%%%%%%%%%%%%%%%%%%%%%%%%%%%%%%%%%%%%%%%%
\begin{figure}[t]
\centering
\includegraphics[height=7.2cm,width=7.2cm]{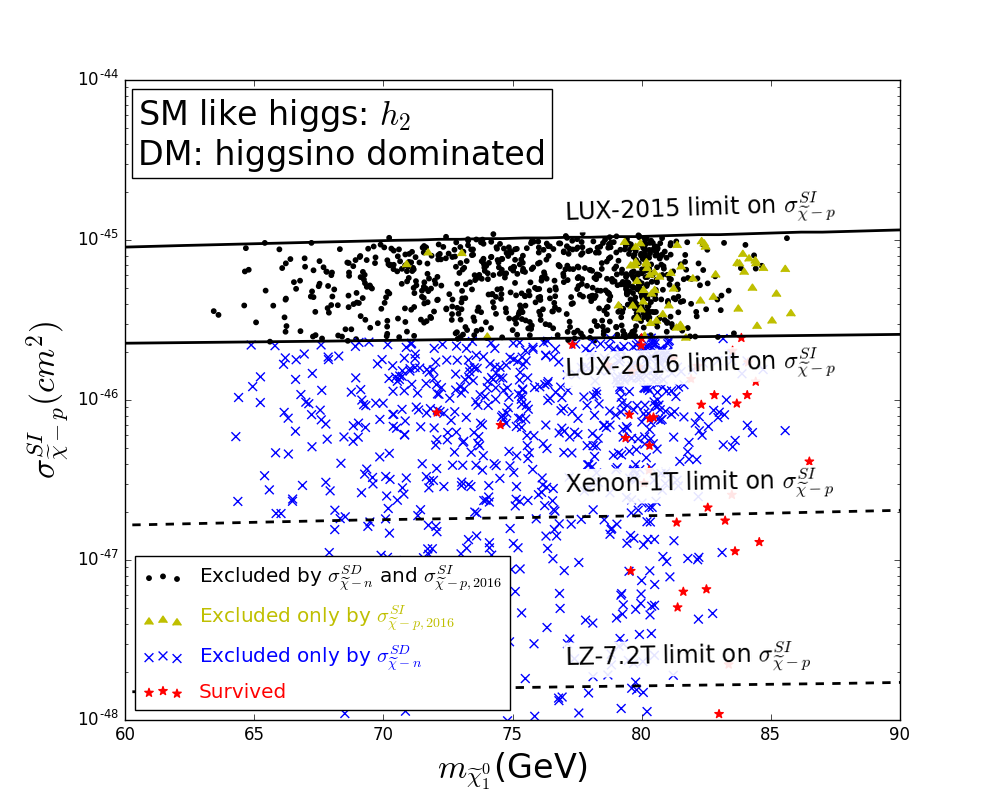} \hspace{-0.5cm}
\includegraphics[height=7.2cm,width=7.2cm]{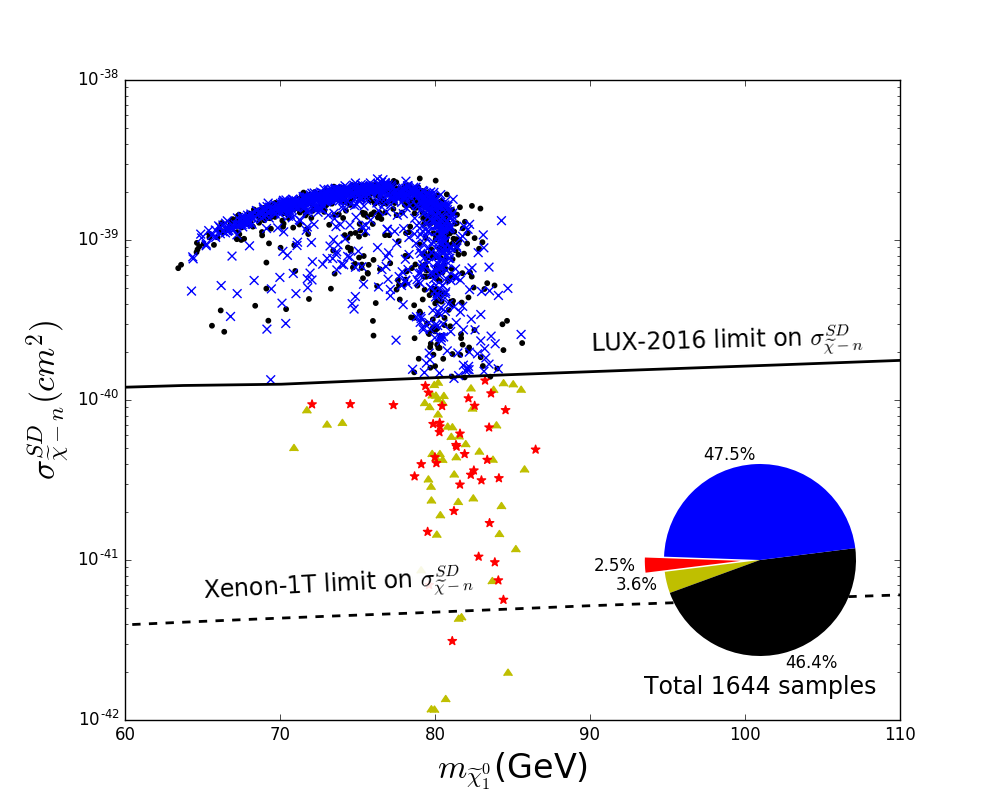}
\vspace{-0.4cm}
\caption{Same as Fig.\ref{fig3}, but for Type IV samples. \label{fig5}}
\end{figure}
%%%%%%%%%%%%%%%%%%%%%%%%%%%%%%%%%%%%%%%%%%%%%%%%%%%%%

In Figs. \ref{fig3}, \ref{fig4} and \ref{fig5}, we project Type II, III and IV samples respectively on
$\sigma_{\tilde{\chi}-p}^{SI}-m_{\tilde{\chi}_1^0}$ and  $\sigma_{\tilde{\chi}-n}^{SD}-m_{\tilde{\chi}_1^0}$ planes.
In these figures the black samples are excluded by both the LUX-2016 limit on SI cross section
and that on SD cross section for DM-neutron scattering. By contrast the dark green ones and the blue ones are
excluded only by either of the limits, and the red ones survive all the limits. The percentages of each kind of colored samples
in the total Type II samples are presented by the disc on the right panel of Fig.\ref{fig3}, and so on for the other
types of samples. Compared with Type I samples, Type II, III and IV samples
exhibit following new features
\begin{itemize}
\item The SI cross sections for the three types of samples may be much lower than the LUX-2016 limit. The underlying reason for such a behavior
is that the singlet dominated $h_1$ is light, and thus it is able to cancel rather strongly the other $h_i$ contributions.
However, one should note that for the more general case without fine parameter structure of blind spots, the SI bounds still provide 
the stronger constraints on the nNMSSM parameter space.
\item  The LUX-2016 limit on SD cross section can be used to exclude the samples obtained in \cite{Cao2016}. Especially for Type III and IV samples,
the exclusion capability becomes so strong that more than $80\%$ of the two type samples are physically forbidden. The reason is as follows.
For Type II samples, the $Z\tilde{\chi}_1^0 \tilde{\chi}_1^0$ coupling
is similar to that for Type I samples, but the parameter $\mu$ for Type II samples may take a significantly lower value (see Table I in \cite{Cao2016}) and thus
the corresponding SD cross section is enhanced. For Type III and IV samples, the explicit expression of the
SD cross section in the heavy gaugino limit is presented in Eq.(\ref{SD-Expression-1}), and it is also due to the relatively
smallness of $\mu$ that the exclusion capability of the SD cross section becomes strong.
\end{itemize}

About our results presented in above figures, we have the following comments:
\begin{itemize}
\item In putting the constraints of the LUX-2016 results on the samples, we directly used the upper bounds on the SI and SD
rates in \cite{LUXCollaboration2016,LUXCollaboration2016-SI} as inputs. However, from experimentalist point of view,
the standard procedure to present the results of a given DM experiment is as follows \cite{DM-detecion-SI-SD}:
First, one of the interaction (SI or SD) is neglected in order to draw conclusions for the other (SD or SI), which is referred to as a pure-SD (resp. -SI) case; then the SI couplings of DM with proton and neutron are assumed to be equal to get the averaged SI limit; finally, for the SD sector the interaction of DM with one type of nucleons
is assumed to dominate and thus the SD results
can be presented in two independent planes: the pure-proton case (equivalently corresponding to
$\langle S_n\rangle = 0$ with $\langle S_n\rangle$ denoting the spin content of neutron in target nuclei)
and the pure-neutron one  (corresponding to $\langle S_p\rangle = 0$).

It is obvious that these assumptions are not
pertinent to our model where both $\sigma^{SI} $ and $\sigma^{SD}_{\tilde{\chi}-p,n}$ are non-zero simultaneously
and the LUX experiment where both proton and neutron contribute to the spin of xenon nuclei.
In this aspect, we note that an improved way to extract constraints from DD experiments was introduced in
\cite{DM-detecion-SI-SD}. From Fig.3 in \cite{DM-detecion-SI-SD} and also the relation
$|\langle S_n\rangle| \gg |\langle S_p\rangle|$ for xenon nuclei, we infer that for most cases,
the upper bounds on $\sigma^{SI}_{\tilde{\chi}-p}$ and $\sigma^{SD}_{\tilde{\chi}-n}$ provided in
\cite{LUXCollaboration2016,LUXCollaboration2016-SI} properly reflect the
capability of the experimental data in constraining theory. This testifies the reasonableness of what
we did in discussing  the constraints of the LUX-2016 results on nNMSSM.

\item We note that the LUX-2016 limit on SD cross section is actually based on the analysis of the experimental data
collected in 2013. Given that the LUX-2016 limit on SI cross section is about 4 times stronger than the corresponding
LUX-2013 result \cite{Akerib:2013tjd,LUXCollaboration2016-SI}, we infer from Eq.(13) of \cite{DM-detecion-SI-SD} that
once the data underlying the LUX-2016 limit
are analyzed for SD cross section, the upper bound on  $\sigma^{SD}_{\tilde{\chi}-n}$ should be lowered
by an approximate factor of 4. If this is true, Type II, III and IV samples in nNMSSM will be further limited.

\item From the monojet analyses performed by ATLAS and CMS collaborations \cite{Monojet-ATLAS,Monojet-CMS},
it seems that the monojet searches at the LHC Run-I are able to put much tighter constraint on DM interactions than the SD limit.
Motivated by this observation, we repeat the analyses of the two collaborations by carrying out detailed simulations,
and surprisingly we do not find any constraint from the monojet searches on nNMSSM (we note that similar conclusion was
obtained in \cite{Schwaller:2013baa}).  The difference, we imagine, originates from the assumption on the mediator of the signal.
To be more specific, ATLAS considered the process
$p p \to j V \to j \tilde{\chi} \tilde{\chi}$ where $\tilde{\chi}$ denotes a DM particle, and
$V$ represents a vector boson with weak couplings to quark and DM pair. As can be seen from Fig.7 in \cite{Monojet-ATLAS},
strong constraint can be set only when $m_V  > 2 m_{\tilde{\chi}}$ so that the vector boson $V$ is actually produced on-shell
and subsequently decays into the DM pair. By contrast, in the NMSSM the monojet signal proceeds
from the process $p p \to j Z^{(\ast)} \to j \tilde{\chi}_1^0 \tilde{\chi}_1^0$. If the $Z$ boson in the process is on shell,
the upper bound from $Z$ invisible decay set by LEP-I experiments has required the monojet plus $E_T^{miss}$ signal sufficiently small,
while if the $Z$ boson is produced off-shell, the $Z$ propagator suppression makes the cross section significantly lower than
that in the ATLAS analysis. As far as the CMS analysis is concerned, it assumed the contact interaction of DM with quarks
(see Fig.5 in \cite{Monojet-CMS}) and consequently some kinematic quantities such as $E_T^{miss}$ distribute in a way
quite different from those of the NMSSM. This will affect greatly the cut efficiencies in the analysis.

\item Since the LUX-2016 results have required the SI cross section to be lower than about $10^{-45} {\rm cm^2}$,
it is quite general that different $h_i$ contributions in Eq.(\ref{SI-cross}) cancel each other to get the
future measurable cross section, which will introduce another fine tuning problem
\cite{DM-fine-tuning-1,DM-fine-tuning-2,DM-fine-tuning-3,DM-fine-tuning-4}.
To investigate this problem, we define the quantity
$\Delta_{DM} = max \left \{ (f_{h_i}^{p}/f^p)^2 \right \}$ to measure the extent of the fine tuning. We find that
among the samples surviving the LUX-2016 constraints, more than $40\%$ of them predict $\Delta_{DM} > 50$. This fact again
reflects the strong limitation of the latest DD experiment on nNMSSM.

\end{itemize}

%%%%%%%%Fig.4%%%%%%%%%%%%%%%%%%%%%%%%%%%%%%%%%%%%%%%%%%%%%%%%%%%%%%%%%%%%
\begin{figure}[t]
\centering
\hspace{0.05cm} \includegraphics[height=7.2cm,width=7.0cm]{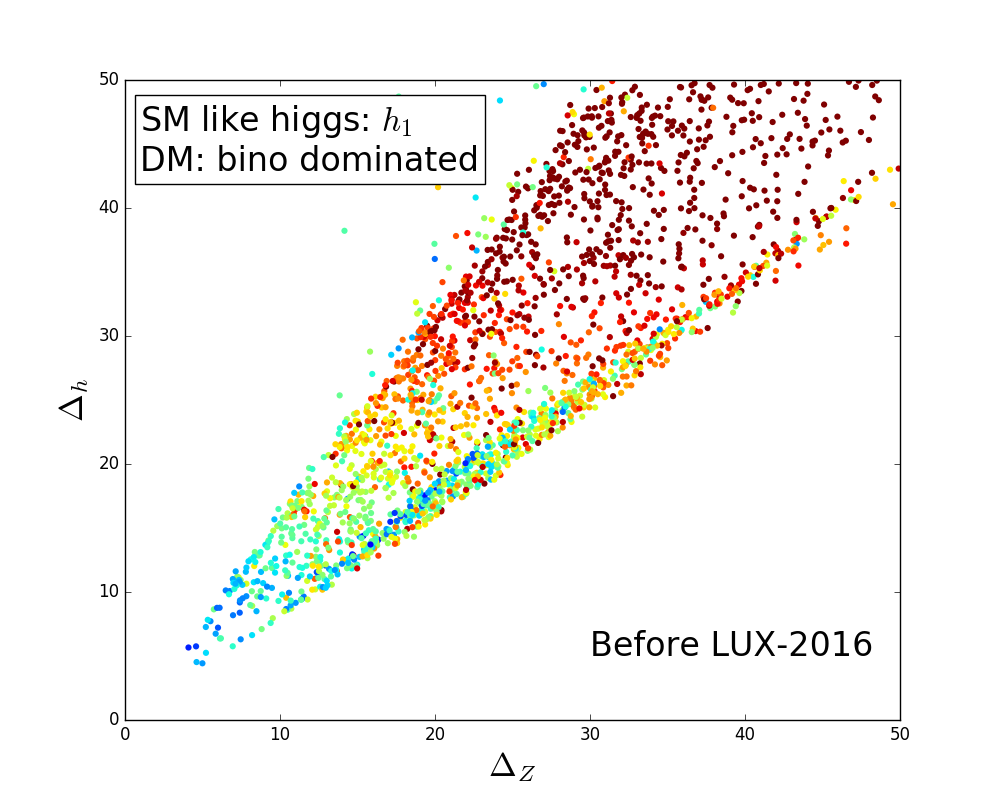} \hspace{-0.7cm}
\includegraphics[height=7.2cm,width=8.4cm]{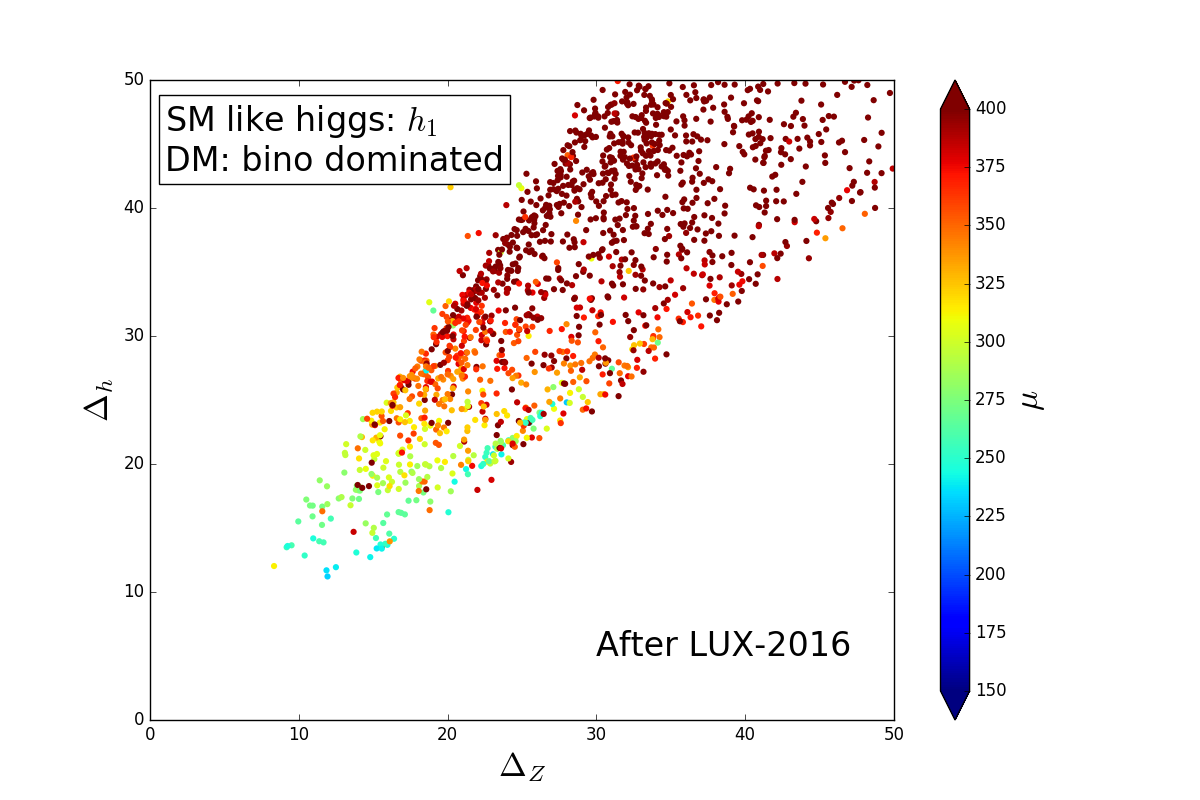}
\vspace{-0.8cm}
\caption{Naturalness of Type I samples in the NMSSM before and after considering the LUX-2016 limits. Different colors represent
the values of $\mu$, which is shown on the right side of the figure.  \label{fig6}}
\end{figure}
%%%%%%%%%%%%%%%%%%%%%%%%%%%%%%%%%%%%%%%%%%%%%%%%%%%%%

%%%%%%%%Fig.4%%%%%%%%%%%%%%%%%%%%%%%%%%%%%%%%%%%%%%%%%%%%%%%%%%%%%%%%%%%%
\begin{figure}[t]
\centering
\hspace{0.05cm} \includegraphics[height=7.2cm,width=7.0cm]{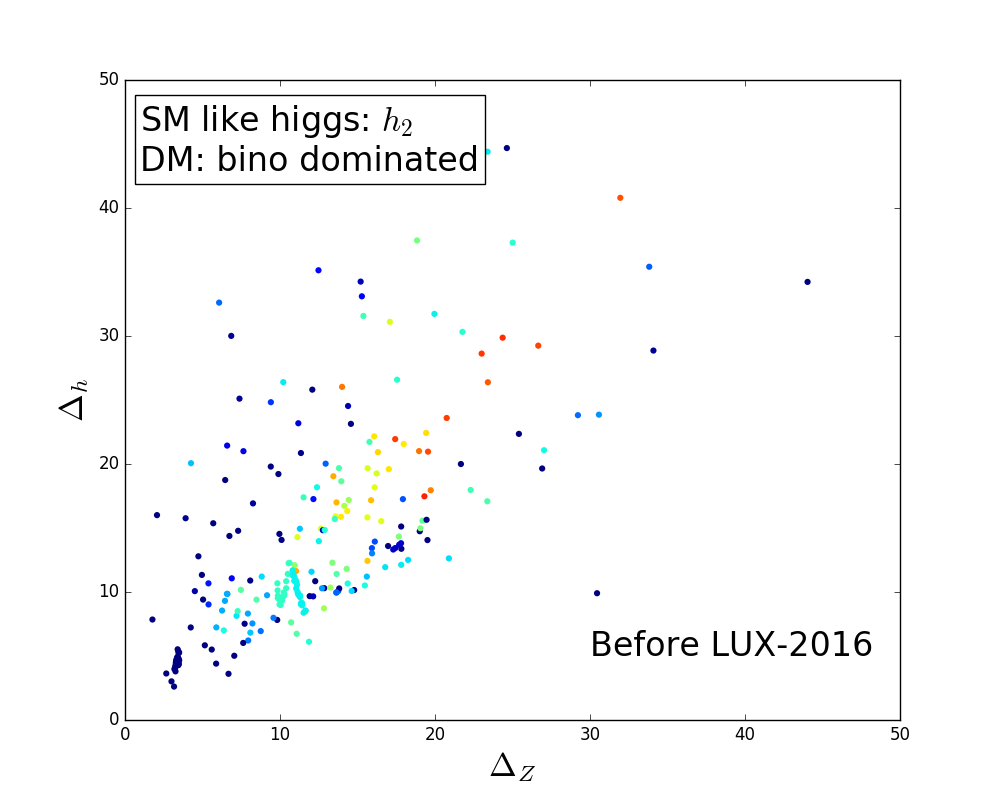} \hspace{-0.7cm}
\includegraphics[height=7.2cm,width=8.4cm]{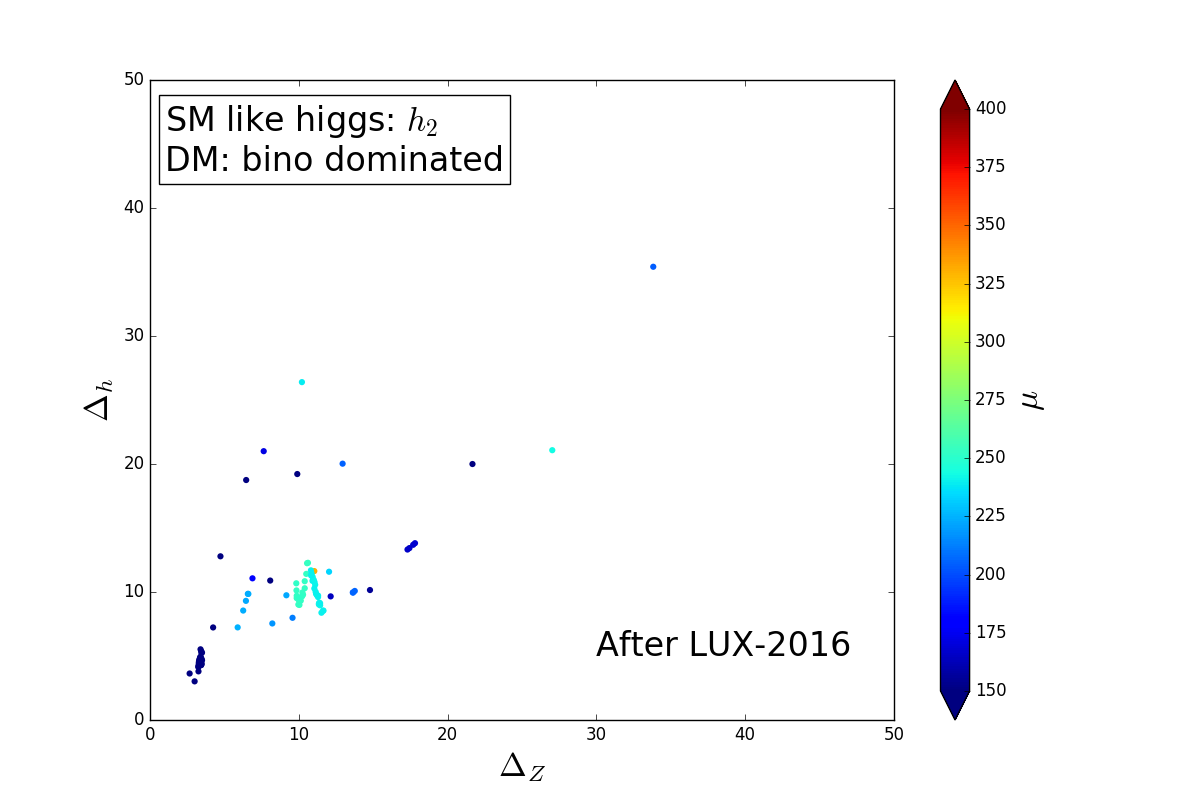}
\vspace{-0.8cm}
\caption{Same as Fig.\ref{fig6}, but for Type II samples. \label{fig7}}
\end{figure}
%%%%%%%%%%%%%%%%%%%%%%%%%%%%%%%%%%%%%%%%%%%%%%%%%%%%%

%%%%%%%%Fig.4%%%%%%%%%%%%%%%%%%%%%%%%%%%%%%%%%%%%%%%%%%%%%%%%%%%%%%%%%%%%
\begin{figure}[t]
\centering
\hspace{0.05cm} \includegraphics[height=7.2cm,width=7.0cm]{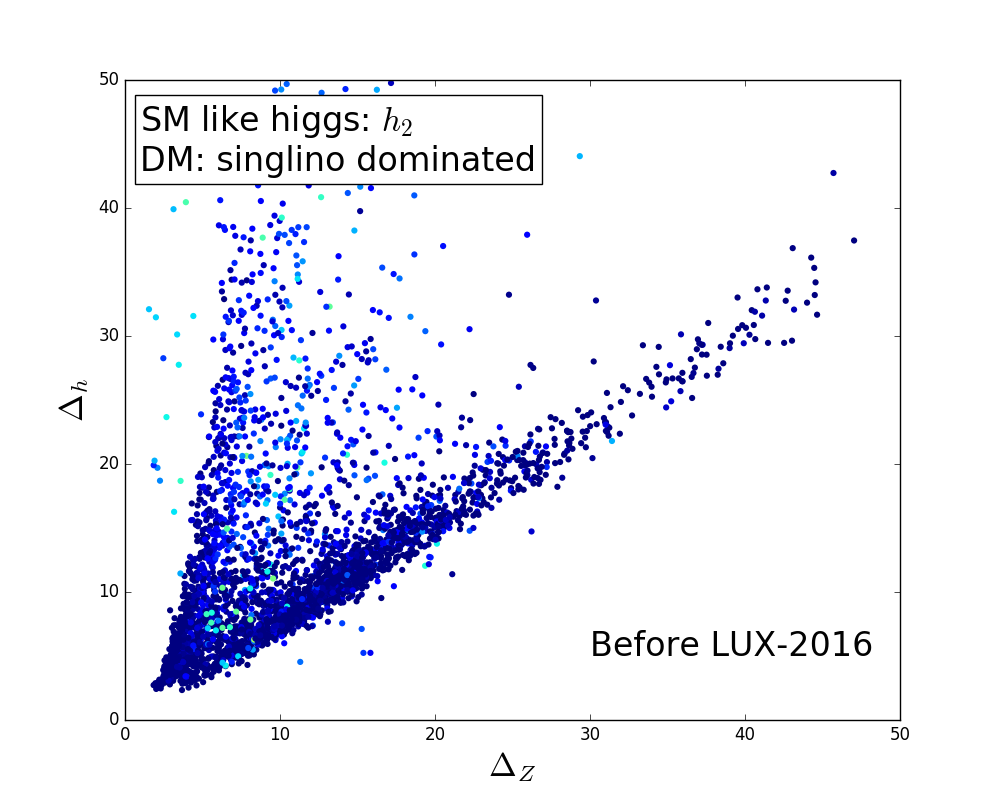} \hspace{-0.7cm}
\includegraphics[height=7.2cm,width=8.4cm]{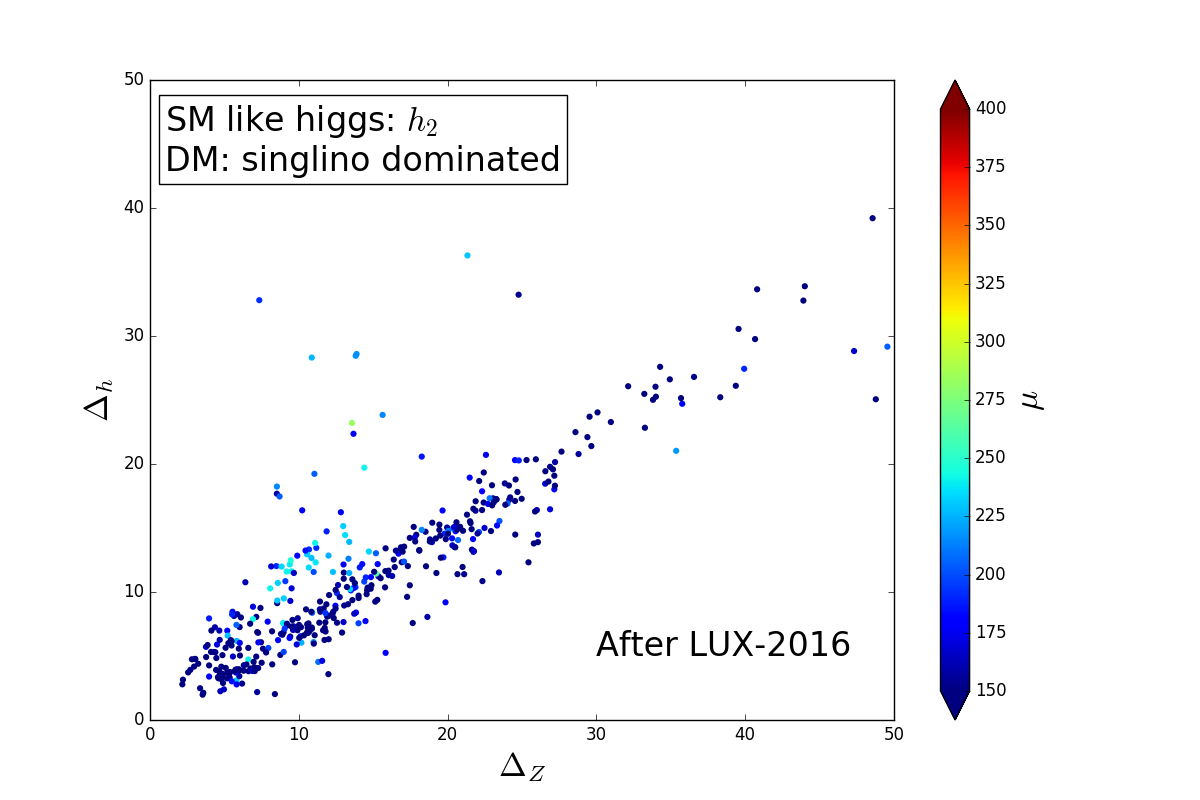}
\vspace{-0.8cm}
\caption{Same as Fig.\ref{fig6}, but for Type III samples.  \label{fig8}}
\end{figure}
%%%%%%%%%%%%%%%%%%%%%%%%%%%%%%%%%%%%%%%%%%%%%%%%%%%%%

%%%%%%%%Fig.4%%%%%%%%%%%%%%%%%%%%%%%%%%%%%%%%%%%%%%%%%%%%%%%%%%%%%%%%%%%%
\begin{figure}[t]
\centering
\hspace{0.05cm} \includegraphics[height=7.2cm,width=7.0cm]{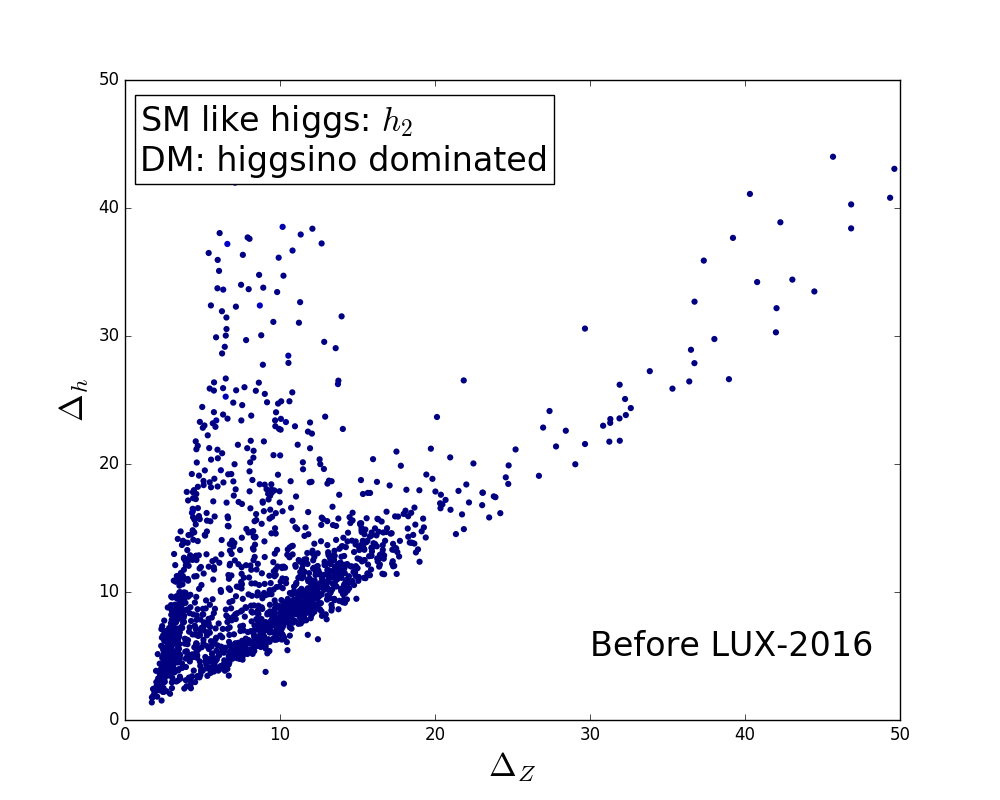} \hspace{-0.7cm}
\includegraphics[height=7.2cm,width=8.4cm]{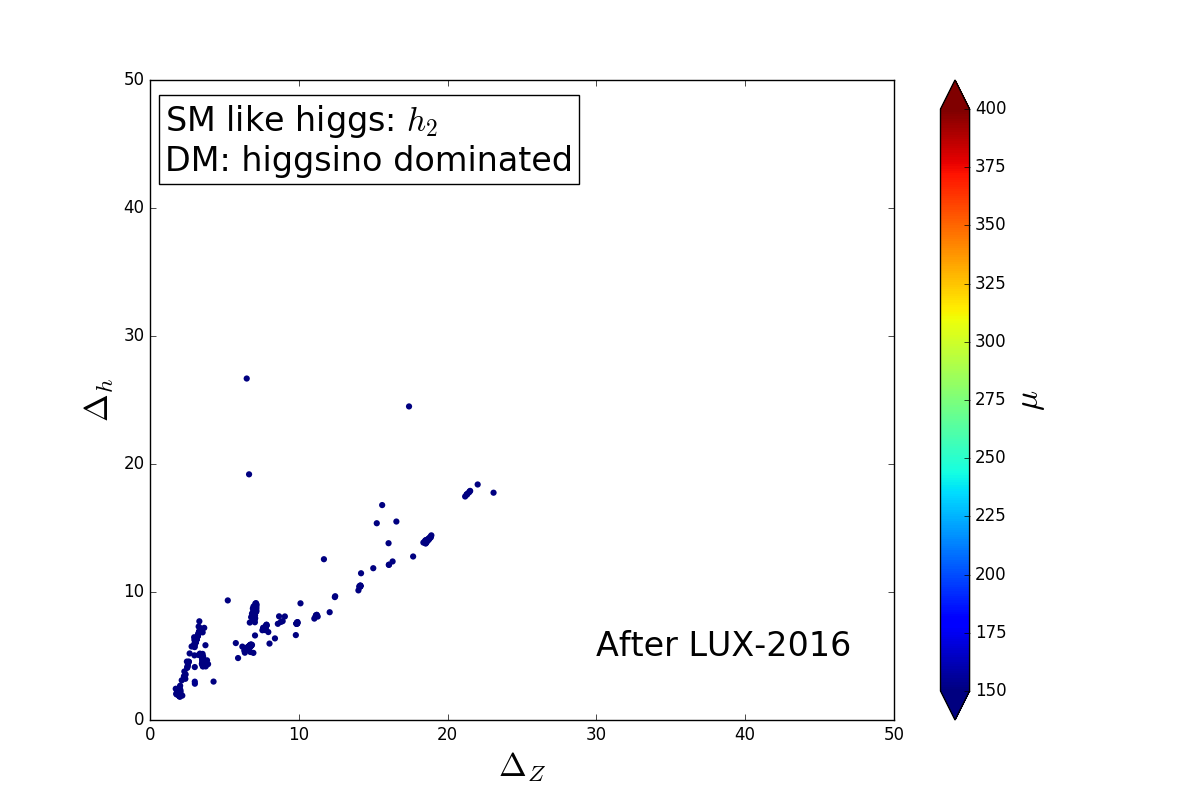}
\vspace{-0.8cm}
\caption{Same as Fig.\ref{fig6}, but for Type IV samples. \label{fig9}}
\end{figure}
%%%%%%%%%%%%%%%%%%%%%%%%%%%%%%%%%%%%%%%%%%%%%%%%%%%%%

%%%%%%%%Fig.4%%%%%%%%%%%%%%%%%%%%%%%%%%%%%%%%%%%%%%%%%%%%%%%%%%%%%%%%%%%%
\begin{figure}[t]
\centering
\includegraphics[height=7.2cm,width=7.2cm]{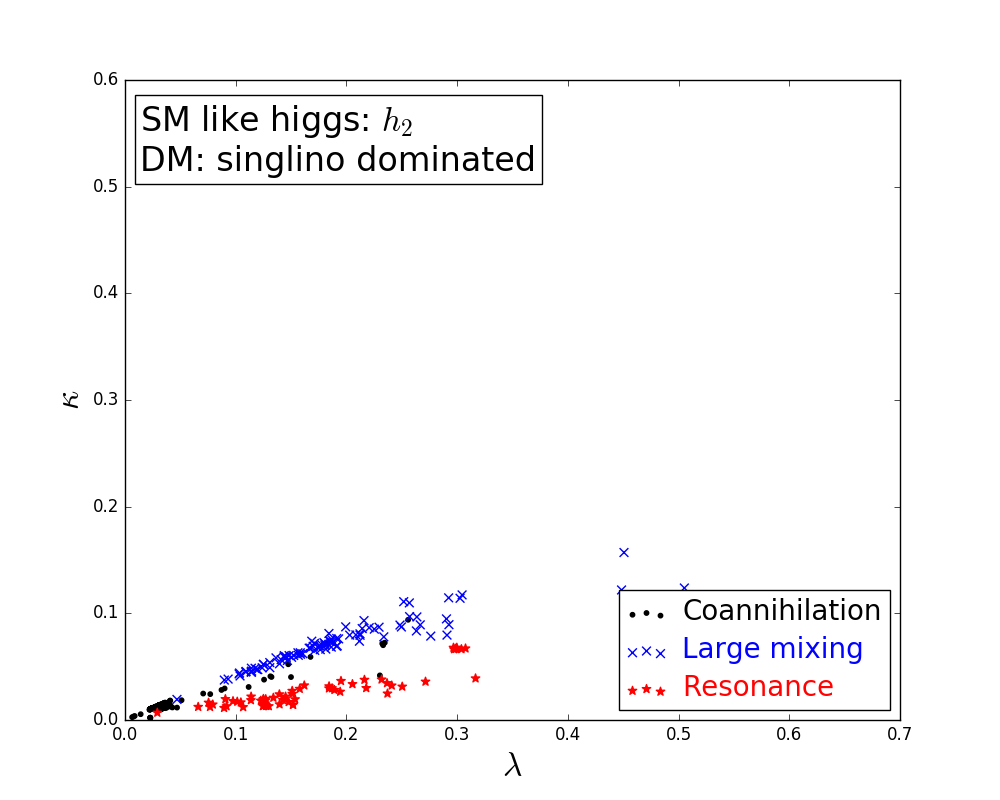} \hspace{-0.5cm}
\includegraphics[height=7.2cm,width=7.2cm]{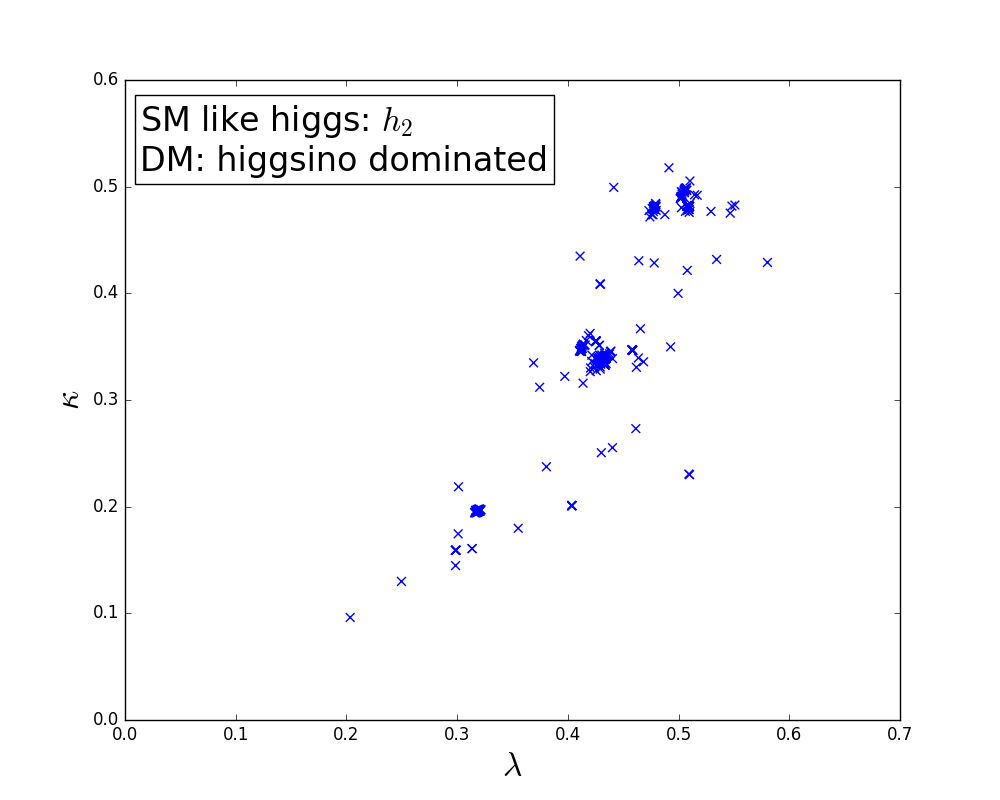}
\vspace{-0.4cm}
\caption{Dominant DM annihilation mechanisms in early universe for Type III (left panel) and Type IV samples (right panel) surviving
the LUX-2016 limits.  \label{fig10}}
\end{figure}
%%%%%%%%%%%%%%%%%%%%%%%%%%%%%%%%%%%%%%%%%%%%%%%%%%%%%

\section{\label{Section-Status}Status of nNMSSM after the LUX-2016 results}

In this section, we investigate the status of nNMSSM after the LUX-2016 results. In order to get
more physical samples for study, we repeat the scan introduced in Section 3 by including the LUX limits
in the likelihood function as what was done in \cite{Likehood}. In the renewed Markov Chain scan,
we find that it becomes rather difficult to get Type II, III and IV samples, and that the surviving
samples are distributed in some isolated parameter islands.
These facts reflect the strong limitation of the LUX-2016 results on nNMSSM.

In Fig.\ref{fig6}, we show the impacts of the LUX-2016 limitations
on $\Delta_Z$ and $\Delta_h$ for Type I samples. Samples on the left panel
are obtained without considering the LUX-2016 constraints,  while those on the right panel satisfy the
constraints. From this figures, one can learn that if the LUX-2016 limits are not considered, $\Delta_Z$ and $\Delta_h$
can be as low as about 5, while the LUX experiment has pushed them up to be larger than 10.  One can also infer that
the experiment has excluded samples with $\mu \lesssim 230 {\rm GeV}$, and by contrast the electroweakino searches at LHC Run-I can not do this. Moreover, we remind from Fig.\ref{fig2} that if the future XENON-1T 
experiment does not detect any DM signal, $\Delta_Z$ and $\Delta_h$ for Type I scenario must be larger than 50, 
and correspondingly the lower bound of the parameter $\mu$ will be larger than $400 {\rm GeV}$.

In Figs. \ref{fig7}, \ref{fig8} and \ref{fig9}, we show similar graphs to Fig. \ref{fig6} for Type II, III and IV
samples respectively. These figures indicate that even after considering the LUX-2016 limits, $\Delta_Z$ and $\Delta_h$ can still
be as low as about 2. This implies that until now the NMSSM is still a viable theory in naturally predicting the quantities.

Since Singlino-dominated or Higgsino-dominated DM is peculiar to the NMSSM, in the following we concentrate on
the Type III and IV samples and study the annihilation of DM in early universe. For this purpose, we project the surviving samples
on the $\lambda-\kappa$ plane with different color representing the dominant annihilation mechanism for each sample.
The corresponding results are shown in Fig.\ref{fig10} with the left panel for Type III samples and the right panel
for Type IV samples. This figure indicates that for the Type III samples, DM may annihilate via the co-annihilation introduced
in section II, large Higgsino and Singlino mixing or resonant funnel to get its measured relic density, while the 
Type IV samples achieve the relic density only through the sizable mixing. This figure also indicates that for the Type III samples,
there are points with $\lambda < 0.1$.  We examined the property of these samples and found that they are characterized by
$v_s > 1\, {\rm TeV}$ while basically all singlet-dominated particles are lighter than about 250 GeV.
Obviously, such a configuration of the Higgs potential is somewhat unnatural since the vacuum expectation value $v_s$ is much larger than the related scalar masses.
Interestingly, we checked the vacuum stability of the potential by the package Vevacious
\cite{Camargo-Molina:2013qva,Camargo-Molina:2014pwa} and found that the vacuums are usually long lived.

\section{\label{Section-Conclusion}Conclusions}

In this work we studied the constraints from the latest LUX results including both SI and SD measurements on natural NMSSM (nNMSSM). Since this theoretical framework
usually favors light Higgsino mass $\mu$ which can induce a sizable rate for the neutralino DM-nucleon scattering, the constraints are expected to be
rather tight. Our main observations include:
\begin{itemize}

\item The SD bound is complementary to the SI bound in limiting nNMSSM since they have different dependence on SUSY parameters.
Especially for the blind spots where the SI cross section vanishes due to strong cancelations among different contributions,
only the SD DM-nucleon scattering contributes to the DM signal in the DD experiments and is therefore
vital for the detection.  We note that since the LUX-2016 experiment has set an upper bound on the SI cross section at about
$10^{-45} {\rm cm^2}$, most samples that survive the bound should lie near the blind spots.

\item For the peculiar scenarios of the NMSSM where the next-to-lightest CP-even Higgs corresponds to the 125 GeV Higgs boson
discovered at the LHC, more than $85\%$ of the samples obtained in our random scan are excluded by the latest LUX results.
Although the exact percentage may vary with different scan strategies, this number can exhibit the high sensitivity of the nNMSSM scenarios to the DD experiment.
By contrast, the constraint from the monojet searches at LHC Run-I on nNMSSM is rather weak due to the small
production cross section, which is constrained by the invisible $Z$ boson width or suppressed by the off-shell $Z$ propagator.

\item Quite distinctively, the SD bound is much more powerful than the SI bound in excluding SUSY parameter for Type III and Type IV samples in our study. This is opposite to the common impression that the SI bound is stronger than the SD one. However, one should also note that for the more general case without fine parameter structure of blind spots, the SI bounds still provide the stronger constraints on the nNMSSM parameter space.

\item Low fine tuning samples are strongly limited within some isolated regions of the NMSSM parameter space and difficult to obtain. Future
dark matter direct search experiments such as XENON-1T will provide a better test of nNMSSM.

\end{itemize}

Finally, we remind that the LUX-2016 limit on SD cross section is actually based on the analysis of the experimental data
collected in 2013. Given that the LUX-2016 limit on SI cross section is about 4 times smaller than the corresponding
LUX-2013 result \cite{Akerib:2013tjd,LUXCollaboration2016-SI}, it is expected that, once the data underlying the LUX-2016 limit
are analyzed for SD cross section, the upper bound on  $\sigma^{SD}_{\tilde{\chi}-n}$ might also be lowered
by an approximate factor of 4. In this case, the constraint from the SD bound will become stronger than
what we obtained in this work.

\section*{Acknowledgement}

Junjie Cao and Peiwen Wu thank  Prof. Yufeng Zhou and Xiaojun Bi for helpful discussion about dark matter direct and indirect detection
experiments. This work is supported by the National Natural Science Foundation of China (NNSFC) under grant No. 11575053 and 11275245.

%\bibliographystyle{unsrt}
%\bibliographystyle{JHEP}
%\bibliography{SD}

\end{document}